\documentclass[prb,twocolumn,aps,showpacs,superscriptaddress,epsfig]{revtex4-1}
\usepackage{amssymb}
\usepackage{amsbsy}
\usepackage{amsmath}
\usepackage{float}
\usepackage{epsfig}
\usepackage{graphicx}
\usepackage{subfigure}
\usepackage[dvipsnames]{xcolor}
\usepackage{soul}
\usepackage{hyperref}
\hypersetup{
	colorlinks=true,
	linkcolor=blue,
	filecolor=cyan,      
	urlcolor=magenta,
	citecolor=red,
}

\newcommand{\andreas}[1]{\textcolor{magenta}{#1}}

\begin{document}

\preprint{APS/123-QED}

\title{Boundary transfer matrix spectrum of measurement-induced transitions}

\author{Abhishek Kumar}
\affiliation{Department of Physics, University of Massachusetts, Amherst, Massachusetts 01003, USA}
\author{Kemal Aziz}
  \affiliation{Department of Physics and Astronomy, Center for Materials Theory,  
Rutgers University, Piscataway, NJ 08854, USA}
\author{Ahana Chakraborty}
  \affiliation{Department of Physics and Astronomy, Center for Materials Theory, Rutgers University, Piscataway, NJ 08854, USA}
\author{Andreas~W.~W.~Ludwig}
\affiliation{Department of Physics, University of California, Santa Barbara, California 93106, USA}
\author{Sarang Gopalakrishnan}
\affiliation{Department of Electrical and Computer Engineering,
Princeton University, Princeton, NJ 08544, USA}
\author{J. H. Pixley}
  \affiliation{Department of Physics and Astronomy, Center for Materials Theory,  
Rutgers University, Piscataway, NJ 08854, USA}
\affiliation{Center for Computational Quantum Physics, Flatiron Institute, 162 5th Avenue, New York, NY 10010, USA}
\author{Romain Vasseur}%
 \email{rvasseur@umass.edu}
\affiliation{Department of Physics, University of Massachusetts, Amherst, Massachusetts 01003, USA}


\date{\today}

\begin{abstract}
Measurement-induced phase transitions (MIPTs) are known to be described by non-unitary conformal field theories (CFTs) whose precise nature remains unknown. Most physical quantities of interest, such as the entanglement features of quantum trajectories, are described by boundary observables in this CFT. We introduce a transfer matrix approach to study the boundary spectrum of this field theory, and consider a variety of boundary conditions. We apply this approach numerically to monitored Haar and Clifford circuits, and to the measurement-only Ising model where the boundary scaling dimensions can be derived analytically. Our transfer matrix approach provides a systematic numerical tool to study the spectrum of MIPTs.

\end{abstract}

\maketitle


\section{\label{sec:level1}Introduction}

Repeated  measurements can drive phase transitions in the entanglement structure of quantum trajectories of many-body quantum systems~\cite{PhysRevB.98.205136,PhysRevB.100.134306,skinner2019measurement,chan2019unitary,PhysRevLett.125.030505,potter2022entanglement,fisher2023random}. The discovery of such measurement-induced phase transitions (MIPTs) has attracted a lot of attention recently~\cite{cao2019entanglement,PhysRevB.100.064204,PhysRevLett.125.030505,PhysRevResearch.2.043072,PhysRevA.102.033316,PhysRevResearch.2.013022,PhysRevB.102.224311,PhysRevB.102.014315,PhysRevB.101.235104,PhysRevLett.125.210602,PhysRevB.102.054302,PhysRevB.102.035119,vijay2020measurement,PhysRevB.103.224210,chen2021non,li2021robust,PhysRevLett.127.235701,PhysRevLett.126.123604,PhysRevB.104.155111,PhysRevLett.126.170602,PRXQuantum.2.010352,PhysRevB.105.064306,sierant2022dissipative,sharma2022measurement,feng2022measurement,iadecola2022dynamical,buchhold2022revealing,PhysRevLett.130.120402,o2022entanglement,PhysRevB.107.224303,Sierant2023entanglement,lemaire2023separate,PhysRevB.108.L041103,jian2023measurement,fava2023nonlinear,ravindranath2023free,PhysRevB.107.214201,PhysRevB.105.094303,tirrito2023full}, mostly in the context of monitored quantum circuits, consisting of entangling unitary gates and disentangling local measurement operators. Focusing on the properties of quantum states conditional on the measurement outcomes, the unitary dynamics results in the  scrambling of quantum information and volume-law entanglement scaling, whereas increasing the rate of local measurements can eventually lead to area-law scaling. This transition can also be equivalently interpreted as a purification transition~\cite{PhysRevX.10.041020}, a quantum coding transition~\cite{PhysRevLett.125.030505,PhysRevB.103.104306,lovas2023quantum}, or a learning transition~\cite{PhysRevX.12.041002,PhysRevLett.129.200602,PhysRevLett.129.120604,majidy2023critical,li2023cross,tikhanovskaya2023universality,ippoliti2023learnability} quantifying how much information the observer learns from the measurement records. 

Given these diverse interpretations and applications, a natural question is to understand the critical behavior of this transition. A crucial step in that direction is provided by exact mappings onto replica statistical mechanical models where the transition is interpreted as an ordering transition from ferromagnetic (volume law) to paramagnetic (area law) phases~\cite{PhysRevB.101.104301,PhysRevB.101.104302,PRXQuantum.2.010352,li2021statistical,li2023entanglement} -- see also~\cite{HaydenRTN,PhysRevB.100.134203,PhysRevB.99.174205} for earlier results on random tensor networks and random unitary circuits. In turn, this statistical mechanics mapping can be used to formulate effective field theory descriptions of MIPTs~\cite{PhysRevB.101.104302,PRXQuantum.2.010352,2023arXiv230307848N}, which remain to be fully understood.
A key prediction of the statistical mechanics mapping is the emergence of conformal invariance at the critical point, which was first observed numerically in monitored Clifford circuits~\cite{PhysRevB.104.104305}. 
More precisely, MIPTs in 1+1d generic monitored quantum systems are described by non-unitary conformal field theories (CFTs) with central charge $c=0$, also known as logarithmic CFTs~\cite{GURARIE1993LogCFTs,GURARIE_LUDWIG2005,Cardy_2013_LCFT_review} (or log-CFTs for short). Consequently, conformal invariance plays a crucial role in precise characterization of the nature of MIPTs~\cite{PhysRevB.104.104305,PhysRevB.101.060301,li2021statistical,PhysRevB.106.214316,PhysRevLett.128.010604,PhysRevLett.128.050602,tikhanovskaya2023universality}.

While a full analytic understanding of the CFTs underlying MIPTs remains out of reach at the moment, it is possible to utilize conformal invariance to study their properties numerically, at least in one dimension.
In 
recent work~\cite{PhysRevLett.128.050602}, it was argued that the quantum evolution with fixed measurement outcomes can be interpreted as a disordered transfer matrix which can be used to extract critical properties using standard CFT tools. This approach was used to extract new universal properties, but also provided numerical estimates of various bulk scaling dimensions accurate enough to distinguish MIPTs in generic monitored circuits (sampled with Haar measure) from those in Clifford monitored circuits.

This transfer matrix approach relies on using periodic boundary conditions. It probes the bulk properties of the underlying CFT---including the effective central charge, the order parameter, and the energy operator scaling dimensions---%
by putting it on an effectively infinitely long cylinder, and applying a finite size scaling analysis in the circumference. 
However, many physical quantities of interest are in fact {\em boundary} observables in the statistical mechanics model and in the CFT. This is because of the nature of the corresponding statistical mechanics model, which is defined on the geometry of the circuit: any physical quantity (including entanglement) computed at a given time $t$ in the monitored quantum circuit will map onto a statistical mechanics observable defined at the top boundary of a two-dimensional lattice (the space-time of the circuit). For example, the von Neumann entanglement entropy of an interval of size $L_A$ scales logarithmically with $L_A$ at criticality (at sufficiently long  times) 
\begin{equation}
    S_A \sim \gamma \ln L_A.
    \label{eqn:SlnL}
\end{equation} 
While in analogy with the entanglement structure of the groundstate of 
translationally invariant, i.e. non-random
CFTs in 1+1d one could naively expect $\gamma$ to be related to the central charge of the underlying CFT~\cite{Calabrese_2009}, this is {\em incorrect}. Instead, the universal coefficient $\gamma$ is related to a boundary scaling dimension -- the scaling dimension of a so-called boundary condition changing (BCC) operator~\cite{PhysRevB.101.104302,PhysRevB.104.104305}, using terminology from boundary CFT (BCFT)~\cite{CARDY1984514,CARDY1989581}.   Note that, while as already mentioned the actual central charge is $c=0$, the quantity playing instead a corresponding role in log-CFTs, including those discussed in this paper, is the so-called  {\it effective central charge} $c_{\rm eff}$, defined in 
Eqs.~(\ref{eqn:3},\ref{DEFceff}) below.

%

In this work, we introduce a boundary transfer matrix approach to study the BCFT data in various families of monitored circuits, which are believed to undergo MIPTs in distinct universality classes. We find that the boundary spectra for distinct universality classes are different. In addition, we numerically evaluate some new boundary exponents that had not yet been computed using different approaches. 


The paper is organized as follows. In section (\ref{sec:level2}) we discuss the boundary transfer matrix approach to extract the boundary spectrum of MIPTs in monitored 1+1d quantum systems. In section (\ref{sec:level3}) we benchmark this approach using the measurement-only Ising model by comparing our numerical results to analytic predictions. In section (\ref{sec:level4}) and (\ref{sec:level5}) we compute the boundary spectra of dual Clifford and dual Haar monitored random circuits, respectively. Finally we conclude in section~(\ref{sec:level6}) by summarizing our results and discussing their broader implications.

\section{\label{sec:level2}Boundary Transfer Matrix}
\begin{figure*}[t!]
	\centering
	\includegraphics[width=0.9\textwidth]{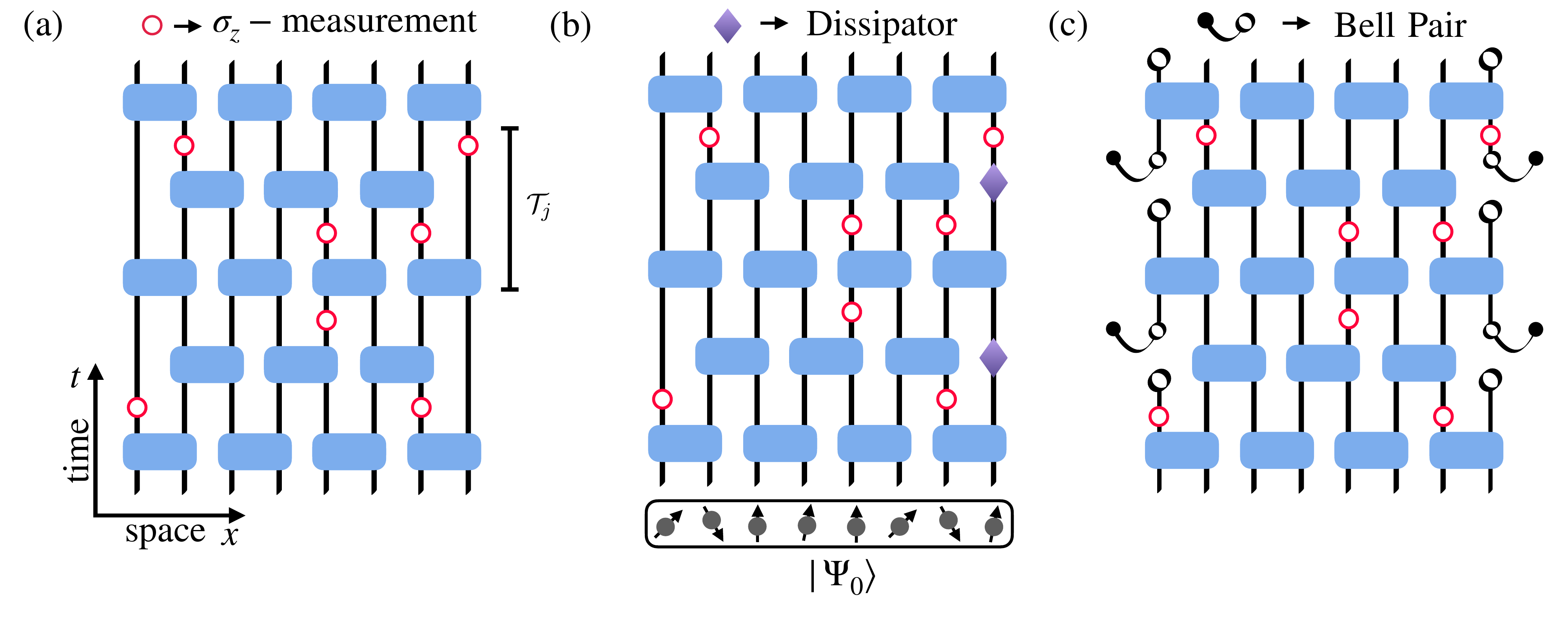}
	\vspace{-2\baselineskip}
	\caption{\textbf{Boundary condition for monitored circuits:}~Implementation of three different boundary conditions for monitored circuits undergoing hybrid dynamics (($a$)--($c$)). It consists of two qubit random unitary gates~(blue rectangles) and $\sigma_{z}$ measurement operations (shown with red circles). All single qubit measurement operations occur at a given space-time location with a probability $p$. 
 The initial state $|\psi_{0}\rangle$ is evolved under hybrid dynamics with boundary ends set to be; (a) open or free, (b) dissipation (shown as purple diamond) at right end and other end is left free, and (c) entangled system-ancilla qubits at both ends. In particular we model dissipation by introducing either dephasing along $z-$axis or by using a maximally mixed depolarizing channel. For the entangled system-ancilla setup we replace the boundary system qubit with a bell pair in each time period where the ancilla set comprise of one qubit from the Bell pairs (denoted by filled black circle) and all the discarded system qubits. The transfer matrix operator which encode the action of unitaries and measurements at the $j$th time step is given by $\mathcal{T}_{j}$.
 }
    \label{fig1}
\end{figure*}

MIPTs are 
most well
studied in hybrid circuits that consist of local projective measurements that are interspersed between two-site random unitary gates arranged in a brick-work pattern~(as shown in Fig~\ref{fig1}). The measurement probability $p$ at a given space-time location is used to drive the transition. The resulting phases exhibit distinct steady state entanglement structure, conditional on measurement outcomes. At low $p$, the dynamics due to entangling unitary gates dominate which in turn results in a subsystem entanglement entropy that scales with the subsystem volume (volume-law), whereas at large $p$ the  local measurements effectively ``collapse'' the many-body wavefunction (area-law scaling). 
At the critical point $p=p_c$, the entanglement scales logarithmically with the subsystem size following Eq.~\eqref{eqn:SlnL} and conformal invariance emerges.

In this work, we probe the boundary conformal properties (BCFT spectrum) of MIPTs by using a boundary transfer matrix, where different boundary conditions can be used to probe the scaling dimension of different BCC operators (i.e. boundary ``observables''). This generalizes the bulk transfer matrix study of Ref.~\onlinecite{PhysRevLett.128.050602}. 

We will apply this approach numerically to monitored circuits that feature Haar and Clifford dynamics, and to  the measurement-only Ising model where the boundary scaling dimensions can be derived analytically.
The Haar and Clifford circuits are realized by drawing the two-qubit unitary gates from the Haar distribution and the finite Clifford group, respectively. However, 
the transfer matrix spectrum
extracted for random circuits with gates drawn from these groups have large error bars due to the non-universal anisotropy factor ($\alpha$) from the asymmetry in the circuit space and time direction~\cite{PhysRevLett.128.050602}. This is improved by restricting the circuit dynamics to a smaller gate set called dual unitary gates~\cite{PhysRevLett.123.210601,PhysRevB.100.064309,PhysRevB.101.094304, PhysRevResearch.4.043212}, which are not expected to change the universality class~\cite{PhysRevLett.128.050602,PhysRevB.101.060301}, but have anisotropy factor $\alpha=1$ since the gates are unitary both in space and time direction.  

As we detail below, we will focus on the following three boundary conditions 
at the two boundary ends of the infinite strip circuit geometry; (i) free boundary conditions, (ii) dissipator boundary using either dephasing or depolarizing channels, and (iii) ``cyclic/swap'' boundary conditions by probing the entanglement properties of boundary ancilla qubits ~(see Fig~\ref{fig1}). This last boundary condition (iii), however, is out of reach with Haar random gates as the number of ancilla qubits must grow with time. 
In spite of this limitation in the special  case (iii), our analysis
allows us to extract various boundary scaling dimensions of various MIPTs, and to further test the emergence of conformal invariance in monitored systems.


\subsection{Transfer matrix}
 
We consider monitored quantum circuits with open spatial boundary conditions,  using various boundary conditions depicted in Fig~\ref{fig1}. To begin, we consider a fixed set of space-time coordinates for the action of measurement operators and unitary gates. The dynamics is then described by the quantum channel  
\begin{equation}
    \mathcal{N}_{t}(\rho) = \sum_{\bf{m}}K_{\bf{m}}\rho K_{\bf{m}}^{\dagger},
    \label{eqn:1}
\end{equation}
where $\rho$ is system's initial density matrix and $K_{\bf{m}} = K_{t}^{m_{t}}K_{t-1}^{m_{t-1}}\dots K_{1}^{m_{1}}$ where 
$K_{s}^{m_{s}} = P_{m_{s}}^{s}U_{s}$ consists of the random unitary operations $U_{s}$ (possibly trivial for measurement-only dynamics) and random projectors $P_{m_{s}}^{s}$ onto the measurement outcomes $m_{s}$. Each term ($K_{\bf{m}}\rho K_{\bf{m}}^{\dagger}$) in the sum of Eq.~\ref{eqn:1} represents a quantum trajectory which occurs with the Born probability $p_{\mathbf m}$ ($=\rm{Tr}(K_{\bf{m}}\rho K_{\bf{m}}^{\dagger})$) for the measurement outcomes $\bf{m}$. The trajectories in the channel form an ensemble of statistical mechanical models with inherent space-time randomness coming from the measurement outcomes. Following Ref.~\cite{PhysRevLett.128.050602}, we introduce the transfer matrix $\mathcal{T}_{j} = K_{2j}^{m_{2j}}K_{2j-1}^{m_{2j-1}}$ (shown in Fig~\ref{fig1}) for the unitary-measurement dynamics which describes evolution for a single time period (maps $\rho \to \mathcal{T}_{j}\rho\mathcal{T}_{j}^{\dagger}$). Now at late times the singular values $\sigma^{\bf{m}}_{i}$ (where $(\sigma^{\bf{m}}_{i})^{2}$ are eigenvalues of $K_{\bf{m}}K_{\bf{m}}^{\dagger}$) of $K_{\bf{m}}$ decay exponentially $\sigma^{\bf{m}}_{i}=e^{\lambda_{i}^{\bf{m}_{t}}}$ (where $\lambda_{i}^{\bf{m}_{t}} < 0$) as the state purifies under evolution of maximally mixed density matrix (where $(\sigma^{\bf{m}}_{i})^{2}$ are its eigenvalues). The trajectory averages of $\lambda_{i}^{\bf{m}_{t}}$ 
give the 
values of the Lyapunov exponents $\lambda_{0},\lambda_{1},\dots$ 
(in descending order)
of the transfer matrix. The leading Lyapunov exponent is related to the average free energy  of the statistical mechanical model up to a factor of time, i.e., $F =-\lambda_{0}t$. Interestingly, this is equivalent~\cite{PhysRevLett.128.050602} 
to the Shannon entropy of the measurement record, where  %
\begin{equation}
    F = -\sum_{\bf{m}}p_{\bf{m}}\ln p_{\bf{m}}.
    \label{eqn:2}
\end{equation}
At criticality, the scaling of this averaged  free energy with system size $L$ is dictated by conformal invariance. This can be seen by introducing the replicated partition functions $\bar{Z}_{k} = \sum_{\bf{m}}(p_{\bf{m}}p_{\bf{m}}^{k})$ from which the 
averaged free energy can be obtained as $F = \lim_{k\to 0 } \frac{dF_{k}}{dk}$ in the replica limit $k \to 0$ where $F_{k}$( $ = -\ln{\bar{Z}_{k}}$) is the replicated free energy. For any finite number of replicas $k$, $F_{k}$ is the free energy of a statistical mechanics model which exhibits 
(for $k$ small enough)
a 2nd order phase transition with emerging conformal invariance~\cite{PhysRevB.101.104301,PhysRevB.101.104302}.
In the replica limit $k\to 0$, this transition coincides with the MIPT. 
To find the free energy density we have to take into account the space-time area ($A$)  where  $A=\alpha L t$ and $\alpha$ is a non-universal anisotropy factor that characterizes the asymmetry between the intrinsic space and time directions of the statistical mechanical model with periodic boundary conditions.
Using standard CFT results~\cite{PhysRevLett.56.746,PhysRevLett.56.742,AFFLECK1988347,francesco2012conformal}, this implies that the bulk free energy density $f(L) = F/(\alpha t L)$  -- scales as\cite{PhysRevLett.128.050602}
\begin{equation}
    f(L) = f(L=\infty) - \frac{\pi c_{\rm{eff}}}{6L^2} + \dots
    \label{eqn:3}
\end{equation}
for a cylindrical geometry of circumference $L$ and circuit depth $t$ (when $t \gg L$), where $f(L=\infty)$ is the extensive bulk term, and the effective central charge 
\begin{equation}
  c_{\rm{eff}} = \lim_{k\to 0} \frac{{ d}c(k)}{{ d}k}, 
  \label{DEFceff}
\end{equation}
is a universal number that characterizes the log-CFT. Note that the actual central charge of the  MIPT CFT is $c=\lim_{k \to 0} c(k) =0$, since the partition function $\bar{Z}_{k}$ becomes trivial $\bar{Z}_{k\to 0}=1$ in the replica limit. Thus the free energy is trivial$(=0)$ with no system size dependence and hence $c=0$. In this bulk geometry, typical values of scaling dimensions are extracted from differences of higher (subleading) Lyapunov exponents, as shown in 
Ref.~\onlinecite{PhysRevLett.128.050602}.


\subsection{Boundary CFT spectrum}

\begin{figure}[t!]
	\centering
 	\vspace{-0.85\baselineskip}
	\includegraphics[width=0.54\textwidth]{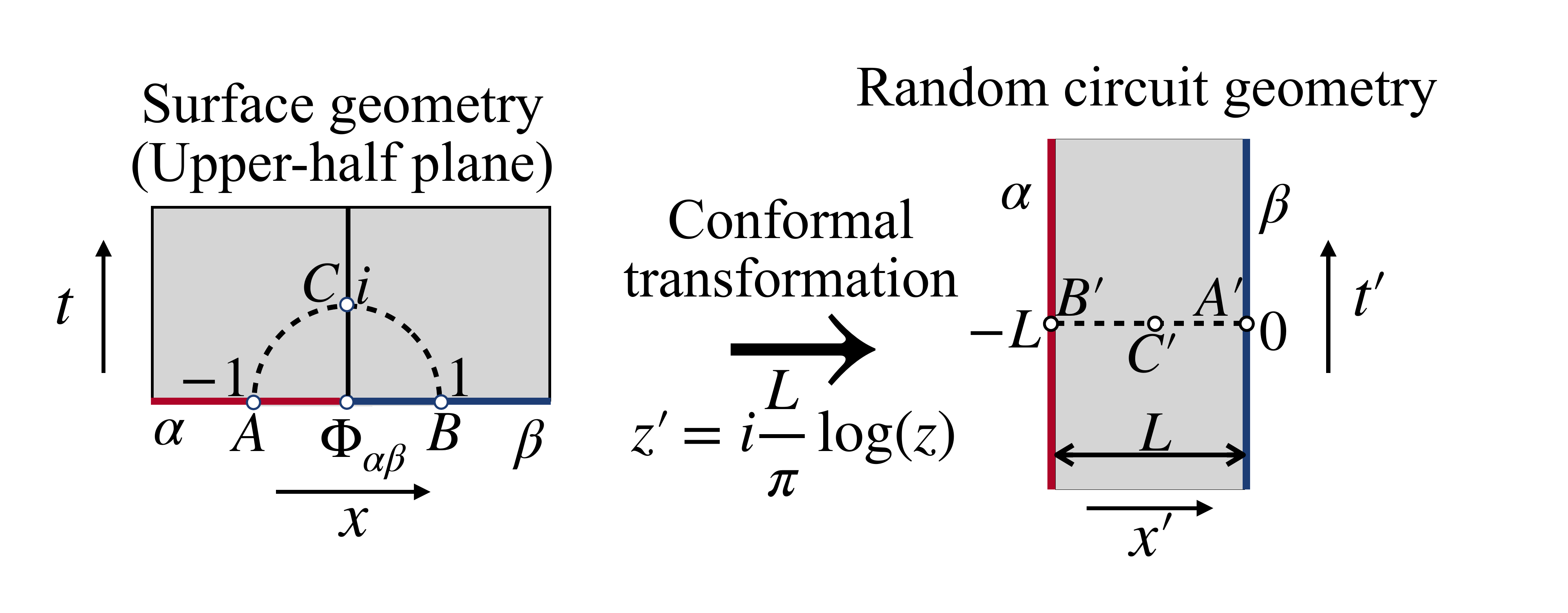}
	\vspace{-2\baselineskip}

	\caption{\textbf{Conformal mapping:} The conformal transformation ($z^{\prime} = \frac{iL}{\pi}\rm{log}(z)$) of upper-half plane maps to infinite strip of width $L$. The two boundary ends $\alpha$ (shown in red) and $\beta$ (shown in blue) of infinite strip geometry results in the insertion of boundary observable $\Phi_{\alpha\beta}$ at the origin of the upper-half plane that separates the boundaries $\alpha$ and $\beta$ in the surface geometry. A semicircle (ACB)  gets mapped to an equal-time line segment joining the boundary strip ($\rm{A}^{\prime}\rm{B}^{\prime}$) as shown by the black dashed line. The respective boundary condition shown in Fig~\ref{fig1} will be denoted by $(\alpha,\beta)= (f,f), (f,a),$ and $(a,b)$, respectively -- using notations consistent with Ref.~\cite{PhysRevB.104.104305}. 
 }
\label{fig2}
\end{figure}

In order to study boundary scaling properties, we now turn to a different geometry, consisting of an infinite strip of width $L$ as shown in Fig~\ref{fig2}. 
This coordinate system $(x',t')$ with $z'=x'+it'$ can be related to the upper half of the complex plane via the conformal transformation $x'+it'=z'=(i L/\pi )\log(z)$ with $z=x+it$. We introduce  distinct boundary conditions on the bottom of the upper half plane, where the left boundary condition is depicted as red and labelled $\alpha$ and the right boundary conditions is blue and labelled as $\beta$ as shown in Fig.~\ref{fig2}, corresponding to the insertion of  BCC operator $\Phi_{\alpha\beta}$ at the origin. Upon applying the conformal transformation, this maps to distinct boundary conditions on the left and right edges of the strip (quantum circuit). In the strip geometry, the BCC operator $\Phi_{\alpha\beta}$ is inserted at imaginary time $\tau =-\infty$, and changes the boundary conditions from $\alpha$ to $\beta$ from the left to the right side of the strip.  As a result of the different (conformally invariant\footnote{Because the bulk is critical, any boundary condition implemented at the microscopic lattice scale of the circuit will at long scales result in a scale-invariant and conformally invariant boundary condition.}) boundary conditions $\alpha$ and $\beta$, the
scaling of 
the averaged free energy density
is given by~\cite{PhysRevLett.56.742,cardy1984conformal}
\begin{eqnarray}
      f_{(\alpha|\beta)} = f(L=\infty) + \frac{f_{{s}}^{(\alpha|\beta)}}{L} +\frac{\pi h_{\alpha|\beta}}{L^{2}} - \frac{\pi c_{\rm{eff}}}{24L^{2}}.
     \label{eqn:4}
 \end{eqnarray}
Compared to~\eqref{eqn:3}, there is an  extra non-universal $1/L$ dependence due to the presence of the surface free energy $f_{{s}}^{(\alpha|\beta)}$, and a boundary specific universal contribution from the scaling dimension $h_{\alpha|\beta}$ ($ = 0$, when $\alpha = \beta$). As illustrated in Fig~\ref{fig2}, this scaling dimension corresponds to the insertion of a BCC operator $\Phi_{\alpha\beta}$ at imaginary time $\tau =-\infty$ which changes boundary conditions from $\alpha$ to $\beta$ from the left to the right side of the strip. 
Just like the effective central charge, the scaling dimension $h_{\alpha|\beta}$ is obtained as a derivative $h_{\alpha|\beta} = \lim_{k\to 0} \frac{dh_{\alpha|\beta}(k)}{dk}$ from the replicated theory (and represents a typical scaling dimension). The surface free energy $(f_{s}^{\alpha | \beta})$ contribution occurs in any statistical mechanics model with specified boundary ends, including for an example an Ising model on a strip.

Using eq.~\eqref{eqn:4}, we see that we can extract numerically the BCC scaling dimensions $h_{\alpha|\beta}$ (belonging to the
boundary spectrum of the CFT) from finite size scaling using various sets of boundary conditions $(\alpha, \beta)$. The boundary conditions that we will consider are guided by the underlying replica statistical mechanics model~\cite{PhysRevB.101.104302}. We will consider both ``free'' boundary conditions (corresponding to open boundary conditions and denoted by $\alpha=f$), and different ``fixed'' boundary conditions, corresponding to fixing the boundary spins of the statistical mechanics model at the boundary. The degrees of freedom (``spins'') of the replicated statistical mechanics model are permutations of the replicas $g \in S_k$, and we will only consider two particular permutations corresponding to dissipation (identity permutation, label $\alpha=a$), and entanglement (cyclic/swap permutation, label $\beta=b$). This terminology is consistent with Ref.~\onlinecite{PhysRevB.104.104305}.

\subsection{Dissipative boundary setup $(\alpha,\beta)=(f,a)$}

We first introduce dissipators at a boundary end to implement the boundary condition shown in Fig~\ref{fig1}(b). A single qubit maximally mixed depolarizing channel or dephasing along the $z$-axis are used to model dissipation. For random Haar circuits, dephasing and depolarizing channels are expected to flow -- in the renormalization group (RG) sense -- to the same conformally invariant boundary condition. These are added after each time period as shown in Fig~\ref{fig1}(b). The resulting dynamics is described by the quantum channel
\begin{equation}
    \mathcal{N}_{t}^{(\mathcal{D})}(\rho) = \mathcal{D}(\mathcal{T}_{\bf{t}}\dots\mathcal{D}(\mathcal{T}_{\bf{2}}(\mathcal{D}(\mathcal{T}_{\bf{1}}\rho \mathcal{T}_{\bf{1}}^{\dagger}))\mathcal{T}_{\bf{2}}^{\dagger})\dots\mathcal{T}_{\bf{t}}^{\dagger}),
    \label{eqn:5}
\end{equation}
where $\mathcal{D}$ is the single qubit depolarizing/dephasing map\cite{nielsen2010quantum} that acts on the boundary qubit $x=L$. The maximally mixed depolarizing channel maps $\rho \to \mathcal{D}(\rho) = I/2$, which models absolute decoherence and the dephasing channel maps  $\rho \to \mathcal{D}(\rho) = P_{\uparrow}\rho P_{\uparrow}+P_{\downarrow}\rho P_{\downarrow}$, which amounts to adding measurements on this qubit and discarding the measurement outcomes~\cite{PhysRevLett.129.080501,li2023statistical}. This random circuit geometry will be denoted by the boundary conditions $(\alpha,\beta) = (f,a)$, and in the statistical mechanics model language it corresponds to fixing boundary spins at the right boundary to the identity permutation.  

Mapping this boundary condition back to the half-plane, we see that the $(f,a)$ boundary condition corresponds to a setup in which one measures the spins on the left half-line at the final time, and computes the resulting Born probability, $\mathrm{Tr}(|\psi\rangle_\mathbf{\tilde{m}} \langle\psi|_\mathbf{\tilde{m}})$, where $\mathbf{\tilde{m}}$ is a bit-string of measurement outcomes for both the mid-circuit measurements in the spacetime bulk of the circuit and the measurements on the left half-system on the final time-step. This quantity receives a nonuniversal surface contribution (which would exist even if all the sites were independent) and a subleading part capturing the long-range correlations between the outcomes of distant measurements.

Another observable that contains direct information about the (typical) scaling dimension $h_{f|a}$ turns out to be the {\it Shannon entropy of the measurement record of the circuit at early time}, i.e. that of a shallow circuit.
To see this consider a ``sideways variant" of Fig.~\ref{fig2}, where the roles of space and time are exchanged: Specifically, it is  convenient to first consider a circuit  with periodic boundary conditions in space of ``large'' circumference $L$, and ``small" depth $t$, so that $ t \ll L$. Then choose as  initial condition  the state labeled by the boundary condition $f$ (which can be represented, e.g., by a simple product state); the state of the circuit appearing at depth $t$, the `final' time,  which contains the physical qubits, is represented by boundary condition $a$
\footnote{For both of these interpretations of the boundary conditions $f$ and $a$, as (simple product) initial state and final state containing the physical qubits, respectively, see Fig. 2 (a), and the corresponding text of Ref.~\onlinecite{PhysRevB.104.104305}.}. Now consider the partition function
$Z_{\mathbf m}$
of the so-defined shallow circuit which equals\footnote{\andreas{paragraph below Eq.~\ref{eqn:1}
}} the Born probability for the measurement record, $p_{\mathbf m}=$~$Z_{\mathbf m}$. 
For this shallow circuit we are considering now, where $t \ll L$, the
corresponding Born-probability-averaged free energy 
density, which  by definition, Eq.~\ref{eqn:2}, is the corresponding shallow-circuit  Shannon entropy density, 
has the same form\footnote{
\andreas{see, e.g., Ref.~\onlinecite{PhysRevB.104.104305}}}
as Eq.~\ref{eqn:4}, except that the roles of space and time are exchanged,
\begin{eqnarray}
\nonumber
&&
- \lim_{L \to \infty}{\overline{
\ln Z_{\mathbf m}}
\over \alpha t L}
=
-
\lim_{L \to \infty}
{1\over \alpha t L }
\sum_{\mathbf m} \ p_{\mathbf m} \ln p_{\mathbf m}=
\\ 
\label{LabelEqShannonEarlyTime}
&&
=
f(t=\infty)  + {f_s^{(f|a)}
\over t}
+ 
{
\pi \ h_{f|a}
\over t^2}
- {\pi \ c_{\rm eff}
\over 24 t^2},
\qquad 
\end{eqnarray}
where $\overline{O}_{\bf m} = \sum_{\bf m} p_{\bf m}  O_{\bf m} $ refers to average over quantum trajectories, and $f(t=\infty)$ and $f_s^{(f|a)}$ are non-universal.
Eq.~\ref{LabelEqShannonEarlyTime} shows that the ${1\over t^2}$-decay of the Shannon entropy 
density at early times is directly affected by the exponent $h_{f|a}$,  a universal signature of the sensitivity of this  entropy to the initial state $f$. At long times, the circuit ``forgets'' about the initial state $f$, and at finite spatial circumference $L$
the behavior in Eq.~\ref{eqn:3} 
obtains in the opposite limit of a deep circuit where $t \gg L$, in the steady state, for the case of periodic spatial boundary conditions we are currently considering. (Open spatial boundary conditions do not modify the early-time $t \ll L$ behavior\cite{PhysRevB.104.104305}, Eq.~\ref{LabelEqShannonEarlyTime},
but the late-time form Eq.~\ref{eqn:4} obtains in the opposite limit of a deep circuit,
$t \gg L$, 
in the steady state.)


\subsection{Boundary ancillas measurement-induced entanglement setup $(\alpha,\beta)=(a,b)$}

In order to implement different boundary conditions, we introduce ancilla qubits at left and right edges~\cite{PhysRevB.104.104305} as shown in Fig.~\ref{fig1}(c). Every time step, we introduce two fresh system qubits that individually form a Bell pair with an ancilla qubit and inject them into the system at the first ($x = 1$) and last ($x=L$) qubit location. Then after the evolution of system qubits for one time period we take out the first and last qubit to store them as 
ancillas with no further action on them. After that we repeat by introducing two fresh pre-entangled qubits at the boundary ends again, and so on. Thus for a circuit of depth $t$, we introduce $2t$ ancilla qubits. Tracing over the ancillas at a boundary
effectively implements dissipation, corresponding to $\alpha=a$. \\

We now compute the entanglement between the right and left ancillas, while measuring 
all 
physical 
qubits -- corresponding to a free $f$ boundary condition on the top layer of the circuit\footnote{\andreas{see Ref.~\onlinecite{PhysRevB.104.104305}, Appendix C, and Ref.~\onlinecite{PhysRevA.71.042306}.}}. This measurement-induced entanglement (MIE)~\cite{PhysRevA.71.042306,Lin2023probingsign} between right and left boundaries effectively implements a change in boundary conditions: one of the
boundaries is traced over (boundary condition $\alpha=a$), while the other is subject to a partial trace (corresponding to a cyclic permutation of replicas, boundary condition $\beta=b$). In the statistical mechanics language, this forces the insertion of a domain wall propagating vertically between the two ends of the strip. The MIE is then directly given by the free energy cost of changing this boundary condition:
\begin{eqnarray}
    \frac{S_{\rm{MIE}}}{\alpha Lt} = f_{(a|b)}-f_{(f|f)}=\frac{\pi h_{a|b}}{L^{2}} + \dots
    \label{eqn:6}
\end{eqnarray}
A ``sideway'' version of this geometry, where the space and the time directions of the circuit were exchanged, was considered in Ref.~\onlinecite{PhysRevB.104.104305}. We thus see that this geometry 
allows us to extract the ``entanglement'' scaling dimension $h_{a|b}$ which controls 
the scaling of the entanglement entropy at criticality. In particular, in previous works $h_{a|b}$ was typically extracted 
directly from the logarithmic scaling of entanglement
entropy
(of the physical qubits) at criticality $S_A \sim 2 h_{a|b} \ln L_A $, which 
exhibits the coefficient $\gamma$ of the logarithm 
in Eq.~\eqref{eqn:SlnL},  
$\gamma=2 h_{a|b}$. 
As already mentioned, we can unfortunately not 
use this boundary condition with Haar random circuits as the Hilbert space 
dimension grows  with time as one adds ancilla qubits as depicted in Fig.~\ref{fig1}.

\subsection{Numerical analysis}

In the rest of this paper, we apply this boundary transfer matrix approach numerically. We average free energies over $10^{5}$--$10^{7}$ sampled quantum trajectories where each trajectory is evolved up to time $t=10L$, after an initial equilibration time ($\sim 4L$). The sampling complexity of quantum trajectories limits us to inspect small system sizes, even with state-of-the-art computing platforms. However we still manage to do larger systems, both for Clifford and Haar, as compared to the previous work with periodic boundary\cite{PhysRevLett.128.050602}. All results are shown at criticality $p=p_{c}$, with $p_{c}^{\rm{MOIM}}= 0.5$, $p_{c}^{\rm{DC}} = 0.205$,  $p_{c}^{\rm{DH}} = 0.14$, and $p_{c}^{C} = 0.1596$, for MOIM, dual Clifford, dual Haar, and  Clifford circuits, respectively~\cite{PhysRevLett.125.070606,PhysRevB.101.060301}.
Note that adding dissipators at boundary does not influence saturation time of free energy density. However it results in additional Monte Carlo sampling of trajectories for Haar circuit as described in Appendix D ( Fig~\ref{fig:Haaraggregate}). This further limits the system sizes and thus results in larger error bars for the universal and non-universal quantities as compared to the Clifford circuit. This also leads to larger error bars for the dissipator boundary as compared to the open boundary condition, even within the Haar circuit. 
The space-time asymmetries
for different monitored circuits are characterized by the anisotropy parameter ($\alpha$) and we extract this from the ratio of space and time correlators~\cite{PhysRevLett.128.050602} which gives $\alpha_{\rm{MOIM}} =\alpha_{\rm{DC}}=\alpha_{\rm{DH}}=1$ for MOIM, dual Clifford, dual Haar circuits (which are all isotropic), and $\alpha_{C} =0.61$ for the Clifford random circuit. All error bars are estimated using a bootstrap analysis where the data is bootstrapped over 1000 samples. 
To improve the estimate of universal quantities we use standard double fitting procedure where we successively remove small system sizes ($L<L_{\rm{min}}$) from the fit which in turn accounts for the leading order correction to the averaged free energy density.  

\section{\label{sec:level3}Measurement-only Ising model}

To demonstrate the validity and accuracy of the boundary transfer matrix spectrum approach to MIPTs, we first 
consider the  measurement-only Ising model (MOIM)~\cite{PhysRevResearch.2.023288,PhysRevB.102.094204} where the conformal spectrum can be derived analytically.

\subsection{Measurement-only dynamics}

 Measurement-only circuits are comprised of different non-commuting measurement operators and are free from random unitary gates~\cite{PhysRevResearch.2.023288,PhysRevB.102.094204,PhysRevX.11.011030,lavasani2021measurement,sang2021measurement,PRXQuantum.2.030313,sriram2022topology}. We consider the measurement-only Ising model (MOIM) which describes measurement-only dynamics using non-commuting competitive measurement operators, $\sigma^x_i$ and $\sigma^{z}_{i}\sigma^{z}_{i+1}$, acting on sites $i\in \{1,2,\dots, L\}$ of a one-dimensional spin-1/2 chain. The dynamics involve the projective measurement operator $M[O]$ that measures the observable $O$, which then maps the state $|\psi\rangle$ onto the eigenspace of $O$, i.e.,
$M[O]|\psi\rangle= \frac{P_{\lambda}|\psi\rangle}{\sqrt{\langle\psi|P_{\lambda}|\psi\rangle}}$, with probability $p_\lambda =\langle\psi|P_{\lambda}|\psi\rangle$ where $\lambda$ is the eigenvalue of the corresponding eigenspace of $O$ and $P_\lambda$ is the projection operator 
onto the eigenspace
of $O$ with eigenvalue $\lambda$. For MOIM, the observable $O$ is either $\sigma^x_i$ or $\sigma^z_i\sigma^z_{i+1}$ and the corresponding projection operator $P_\lambda$ is either $\frac{I+ \lambda\sigma^{x}_{i}}{2} $ or $\frac{I+ \lambda\sigma^{z}_{i}\sigma^{z}_{i+1}}{2} $ where $\lambda$ takes value $\pm 1$. 
In each discrete time step, the measurement operators $M[\sigma^{z}_{i}\sigma^{z}_{i+1}]$ ($\equiv M_{e}^{zz}$) and $M[\sigma^x_i](\equiv M_i^x)$ acts randomly on each edge $e = (i,i+1)$ and site $i$ with a probability $1-p$ and $p$, respectively. The layer of $M_{e}^{zz}$ measurements precedes all probable $M_i^x$ operators. The initial state is set to $|\psi_{0}\rangle = |++\dots+\rangle$ where $|+\rangle = \frac{|0\rangle + |1\rangle}{\sqrt{2}}$ is the eigenstate of $\sigma^{x}$ operator and thus the resulting dynamics leaves the state invariant under the operator $\mathcal{C} = \prod_{i=1}^{L} \sigma_{i}^{x}$ which describes the symmetry of this model. The transfer matrix evolving the system for a single time step then acts as 
 \begin{eqnarray}
     |\psi(t+1)\rangle  = M_{\bf{i}}^{x} \otimes M^{zz}_{\bf{e}} |\psi(t)\rangle,
 \label{eqn:9}
 \end{eqnarray}
 where $\bf{e}$ and $\bf{i}$ are 
 the sets of all edges and sites, respectively, on which
the measurement operations ($M_{e}^{zz}$ and $M_{i}^{x}$) act. It has been shown that the model 
exhibits
an entanglement phase transition between two area law phases and the physics at the critical point is described by bond percolation (which is a CFT with central charge $c=0$, as it should to describe a MIPT).

\subsection{Percolation mapping and replicated statistical mechanics model}

\begin{figure}[t!]
	\centering
	\includegraphics[width=0.48\textwidth]{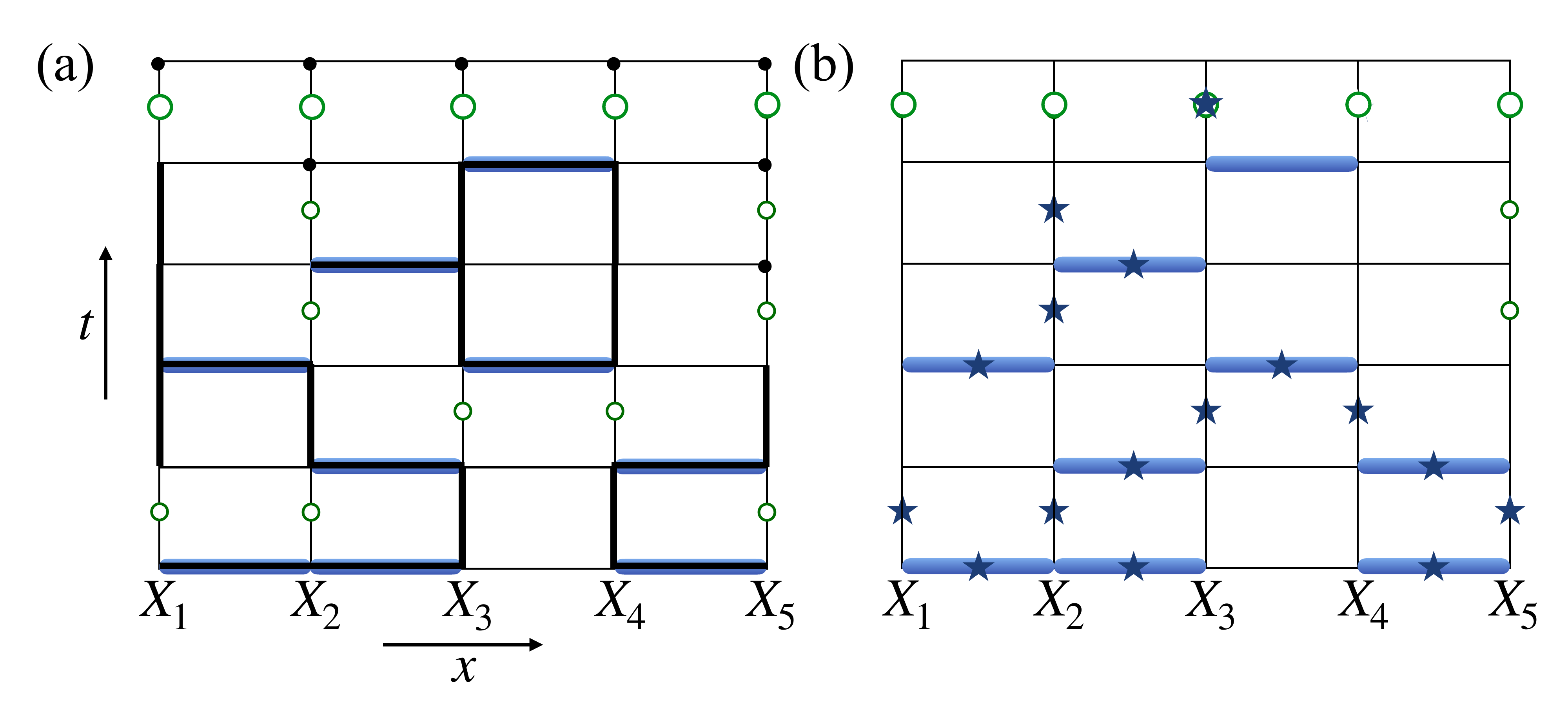}
	\vspace{-1\baselineskip}
	\caption{\textbf{Percolation mapping of the MOIM:} (a) Each $\sigma^{z}_{i}\sigma^{z}_{i+1}$ measurement (blue edge) along an edge maps to a horizontal bond and the absence of a $\sigma^{x}_{i}$ (green circles) measurement corresponds to a vertical bond. This defines a percolation cluster (shown by dark black line). A given time step comprises of a layer of $\sigma^{z}_{i}\sigma^{z}_{i+1}$ and then $\sigma^{x}_{i}$ measurement operations with probability $(1-p)$ and $p$ respectively, that evolves the initial state $(X_{1},X_{2},X_{3},X_{4},X_{5}) $ from $t = 0$ to time $t=5\rm{,~where~} X\equiv \sigma^{x}$. At the last time step, one artificially measures all sites using $\sigma^{x}$ measurement operators. The resulting configuration has parameter $N_{s}= 30$ (number of space-time sites), $N_{\rm{vb}} =11 $ (number of vertical bonds), and  $N_{c} = 11$ (number of percolation clusters, including single-site clusters denoted by isolated black dots). (b) In this example we fix the order of measurement from left to right end in each time step. The total number of random measurements $N_{\rm{rand}}=16$ (marked by stars) is independent of the choice of order in which the measurements are performed. All measured links (vertical or horizontal) which are not marked by a star in this panel correspond to deterministic measurements (meaning the Born probability of that particular measurement outcome is $1$). This example satisfies the relation between percolation clusters and number of random measurements stated in Eq.~\ref{eqn:9}.  }
	\label{fig5}
\end{figure}

 We now review the mapping of this model onto bond percolation~\cite{PhysRevResearch.2.023288,PhysRevB.102.094204}, and generalize it to fully characterize the associated replicated statistical mechanics model $Z_k = \sum_{{\bf m}} p_{\bf m}^{k+1}$. First, note that
to each realization of the circuit,
 one can associate a bond percolation configuration as shown in Fig.~\ref{fig5}(a), where a measurement along an edge $M_{e}^{zz}$ corresponds to a horizontal bond along that edge, whereas the absence of a local measurement $M_{i}^{x}$ on a site maps to a vertical bond. This provides a one-to-one correspondence between percolation configurations and 
 measurement locations. As we now show, the free energy is entirely given by the properties of this percolation configuration. \\

First, note that the MOIM is a stabilizer (Clifford) circuit, so that each measurement outcome is either fully deterministic (determined outcomes), or fully random (with equally probable outcomes). For example let us suppose we have a stabilizer state $|+,+\rangle$. Then measuring $\sigma^{z}_{1}\sigma^{z}_{2}$ will result in either $(|00\rangle + |11\rangle)/\sqrt{2}$ or $(|01\rangle + |10\rangle)/\sqrt{2}$, each with probability $1/2$. However if we measure $\sigma^{x}_{1}$ for the same state, we get back $|+,+\rangle$ with probability 1. We call the former as random and the latter to be  a deterministic measurement outcome. Note that for MOIM whether an outcome is random or deterministic will depend on the order in which measurements are performed in a given layer of time but the total number of random measurements is independent of the choice of order. We choose the order of measurements from left to right end in each time step. As a result, the Born probability of a given trajectory is given by $p_{\bf m} = \left(\frac{1}{2} \right)^{N_{\rm rand}}$, where $N_{\rm rand}$ is the total number of random (non-deterministic) measurements. This means that the free energy is given by $F = \overline{N}_{\rm rand}\ln (2)$, where $\overline{N}_{\rm rand}$ is the average number of random measurements in those circuits. 


\begin{figure*}[t!]
	\centering
	\includegraphics[width=\textwidth]{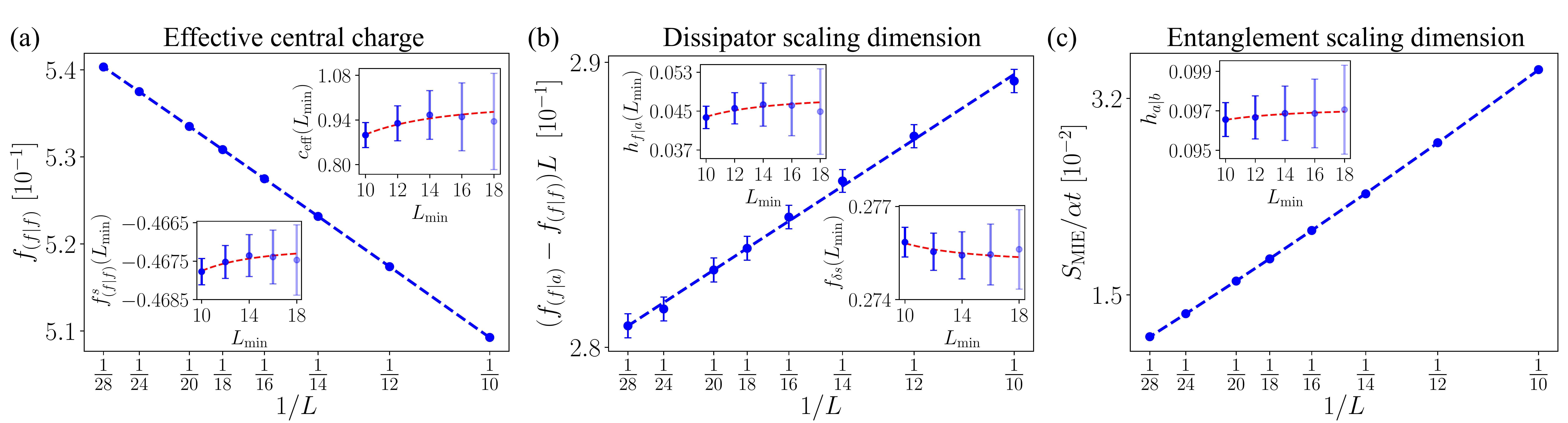}
	\caption{\textbf{Boundary spectrum of MOIM:} We implement the boundary condition in Fig~\ref{fig1} to obtain the respective boundary spectrum of MOIM. (a) The plot of $f_{(f|f)}$ and $1/L$ clearly respect the scaling form stated in Eq.~\ref{eqn:4} (for $L_{\rm{min}} \in \{10,12,14,16,18\}$) as shown by the dashed blue line. In the inset, we perform the double fitting procedure to extract effective central charge $c_{\rm{eff}}$ and free energy surface term $f_{(f|f)}^{s} $, by successively removing the smallest system size and then fitting the data with $L \geq L_{\rm{min}}$. This gives us $c_{\rm{eff}} = 1.0(1)$ and $f_{(f|f)}^{s} = -0.4671(7)$ as shown by the red dashed line. (b) The plot of $(f_{(f|a)}-f_{(f|f)})L$ scales linearly with $1/L$ as shown by the dashed blue line, and from fitting the values of slope and intercept, we obtain the dissipator scaling dimension $h_{f|a} = 0.048(6)$ and $f_{\delta s} = 0.27$, as shown with their respective red dashed lines in the inset plots.  (c) The plot of $S_{\textrm{MIE}}$ vs $1/L$ clearly follows the trend stated in Eq.~\ref{eqn:6} and we find the entanglement scaling dimension $h_{a|b} = 0.097(2)$ using the red dashed line in the inset plot. The fit of $S_{\textrm{MIE}}$ include a leading correction term coefficient $A_{1}= 0.37$. All the data shown above for the MOIM have $p_{c} = 0.5$, $\alpha = 1$, and $L \in \{10,12,14,16,18,20,24,28\}$. We sample $=5\times 10^{6}$ trajectories to find the average value of free energy density.}
	\label{fig3}
\end{figure*}
Next, we notice that the number of random measurements is purely determined in terms of the geometrical properties of the percolation clusters. Namely, we find by inspection that
\begin{eqnarray}
    N_{\rm{rand}} = 2(N_{s}- N_{\rm{vb}} - N_{c}),
    \label{eqn:9}
 \end{eqnarray}
 where $N_{s}$, $N_{\rm{vb}}$, and $N_c$ are the total number of sites, vertical bonds, and disconnected clusters in the percolation configuration. An example illustrating the validity of this expression is provided in Fig.~\ref{fig5} where we fix the order of measurements from left to right end in each time step. For this particular choice, the random and deterministic measurement locations are shown in Fig.~\ref{fig5}(b). So for a fixed measurement trajectory with all measurement outcomes set to $1$, we write the evolution of stabilizer generators for one time step from $t=2$ to $t=3$ in Fig.~\ref{fig5}(b), following the convention $X \equiv \sigma^{x}$ and $Z \equiv \sigma^{z}$ to denote these generators. The stabilizer generators at $t= 2$ are given by $\{X_{1},X_{2},X_{3},X_{4},X_{5}\} $; this set evolves to $\{ X_{1}X_{2}, Z_{1}Z_{2}, X_{3}X_{4}, Z_{3}Z_{4}, X_{5}\}$ upon measuring $\sigma^{z}_{1}\sigma^{z}_{2}$ and $\sigma^{z}_{3}\sigma^{z}_{4}$. Clearly both  measurements are random in accordance with the two site example discussed earlier. Next we measure $\sigma^{x}_{2}$ and $\sigma^{x}_{4}$, which results in the stabilizer generators $\{X_{1}, X_{2}, X_{3}X_{4}, Z_{3}Z_{4}, X_{5}\}$, at the end of time $t=3$. The measurement on site $2$ is random while the measurement on last qubit is deterministic since the qubit already is in the $|+\rangle$ state. Doing this for all time steps in turn leads to the total random measurements that obey the formula in Eq.~\ref{eqn:9}. Although we do not have a formal proof of this formula, we have checked its validity numerically for very large circuits. This equation can further be broken down into purely extensive contributions, and terms including non-extensive universal corrections. The first two terms  scale extensively ($O(Lt)$) and will therefore contribute to the bulk free energy only. Dropping these non-universal extensive contributions, we thus find that 
\begin{equation}
    F \sim  - 2 \overline{N}_{c} \ln 2,
    \label{eqnFpercolation}
 \end{equation}
 where $\overline{N}_{c}$ is the average number of percolation clusters at criticality. This quantity can be computed using the standard 
 Fortuin-Kasteleyn (FK)~\cite{newman1994disordered} mapping between percolation and the $Q\to 1$ limit of the $Q$-state Potts model. Using these standard percolation results, we find that  $\overline{N}_{c} = - \left. \frac{d}{dQ}F^{\rm{Potts}} \right|_{Q=1}$ where $F^{\rm{Potts}}$ is the free energy of the classical $Q$-state Potts model. This provides a direct relation between the free energy of the circuit (Shannon entropy of measurement record) and that of the classical Potts model. In fact, using the same reasoning as above, one can show that the replicated partition function $Z_k = \sum_{{\bf m}} p_{\bf m}^{k+1}$ maps onto a Potts model with $Q=4^{k}$ (up to non-universal contributions), with $Q \to 1$ in the replica limit $k \to 0$. We can then use standard CFT results for the Potts model to infer universal quantities for the MOIM transition. For example,  the central charge of the Potts model~\cite{francesco2012conformal} is given by $c^{\rm{Potts}} = 1 - \frac{6}{m(m+1)}$ where $m= \frac{\pi}{\arccos(\frac{\sqrt{Q}}{2})} - 1$. We thus find that the effective central charge is given by  
\begin{equation}
  \left. c_{\rm{eff}} = \ln 4 \frac{d}{dQ}c^{\rm{Potts}}\right|_{Q=1} = \frac{5 \sqrt{3} \ln 2}{2 \pi} \simeq 0.96. 
  \label{ceffTheory}
 \end{equation}
Scaling dimensions can be identified through this mapping as well, and fit into the ``Kac table''~\cite{francesco2012conformal} $h_{r,s} = ((r (m+1) -s m)^2 -1 )/(4 m (m+1))$.

 
 We now turn to a numerical analysis of the MOIM in the various geometries summarized in Fig. \ref{fig1}. We will first implement the open (free) boundary conditions, 
     Fig~\ref{fig1}(a)),
 to extract  $c_{\rm{eff}}$. Then we introduce dissipation at one end of the boundary to evaluate the dissipator scaling dimension ($h_{f|a}$). Finally to extract the entanglement scaling dimension $h_{a|b}$, we implement the entangled system-ancilla setup (as described in fig~\ref{fig1}(c)). We make use of stabilizer formalism~\cite{PhysRevA.70.052328,gottesman1998heisenberg,PhysRevX.7.031016,PhysRevB.100.134306,stefan_krastanov_2022_7110286}, which allows for efficient classical simulation of large quantum circuits.

\subsection{\label{sec:level3a}Effective central charge ($c_{\rm{eff}}$)}
We first compute the numerical value of the effective central charge in the cylindrical geometry setup with periodic boundary conditions. The plot of free energy density $(f)$, shown in Appendix~\ref{sec:levelB} (Fig.~\ref{fig8}(a)), follows a straight line when plotted against $1/L^2$ as expected from Eq~\ref{eqn:3} and the slope gives $c_{\rm{eff}} = 0.96(1)$. This is in good agreement with the analytic prediction in Eq~\eqref{ceffTheory}. We also extracted a precise value of $f(L=\infty)= 0.55718(2)$, using the linear double fitting procedure. We now implement MOIM with free boundary ends as shown in Fig~\ref{fig1}(a), corresponding to $(\alpha,\beta) = (f,f)$. The free energy density scales in accordance with Eq.~\ref{eqn:4} where $h_{f|f} = 0$ (as expected\cite{PhysRevB.104.104305}) and the resulting fit gives $c_{\rm{eff}} = 1.0(1)$ as shown in Fig~\ref{fig3}(a). 
Note that this estimate is less accurate than the periodic boundary conditions case, largely because of the non-universal surface term $f_{(f|f)}^{s}$ which adds a fitting parameter. In order to reduce the number of fitting parameters, we use the extensive free energy term $f(L=\infty)$ obtained from periodic boundary conditions in Eq.~\ref{eqn:4}. 

\subsection{\label{sec:level3b}Dissipator scaling dimension ($h_{f|a}$)}
Next, we implement boundary depolarizing/dephasing channels as shown in fig~\ref{fig1}(b), to extract the dissipator scaling dimension $h_{f|a}$. In order to get rid of the extensive bulk contribution, we plot the free energy difference
 \begin{equation}
     f_{(f|a)}-f_{(f|f)} = \frac{f_{\delta s}}{L} + \frac{\pi h_{f|a}}{L^2},
     \label{eqn:fdiff}
 \end{equation} 
 where $f_{\delta s} = f^{s}_{(f|a)}-f^{s}_{(f|f)}$. The plot of $(f_{(f|a)}-f_{(f|f}))L$ vs $\frac{1}{L}$ displays a linear trend with $1/L$ as shown in \ref{fig3}(b). Now using the double fitting procedure we find the boundary scaling dimension $h_{f|a}(L) = 0.048 - \frac{0.432}{L^2}$ and the non-universal free energy surface term 
 $f_{\delta s} = 0.27+\frac{0.06}{L^2}$.
We conclude that $h_{f|a} = 0.048(6)$ and $f_{\delta s} = 0.275(1)$.
 Note that the addition of dissipators still preserves the global symmetry $\mathcal{C}$. We also observe that the scaling dimension remains unchanged with the type of dissipators since dephasing along $z-$axis is equivalent to depolarizing in this model. 

 The mapping onto bond percolation can be generalized in the presence of boundary dissipators. We find that boundary dissipation induces additional random (non-deterministic) measurements that are associated with the hulls of percolation clusters touching the boundary, see Appendix~\ref{sec:levelC}. The BCC operator associated with changing the fugacity of such boundary hulls was identified in Ref.~\cite{JACOBSEN2008137}. Combining these results, we find that $h_{f|a} = \frac{\sqrt{3}}{8 \pi} \ln 2 \simeq 0.048$ (see Appendix~\ref{sec:levelC}), in very good agreement with our numerical results.

\begin{figure*}[t!]
	\centering
	\includegraphics[width=\textwidth]{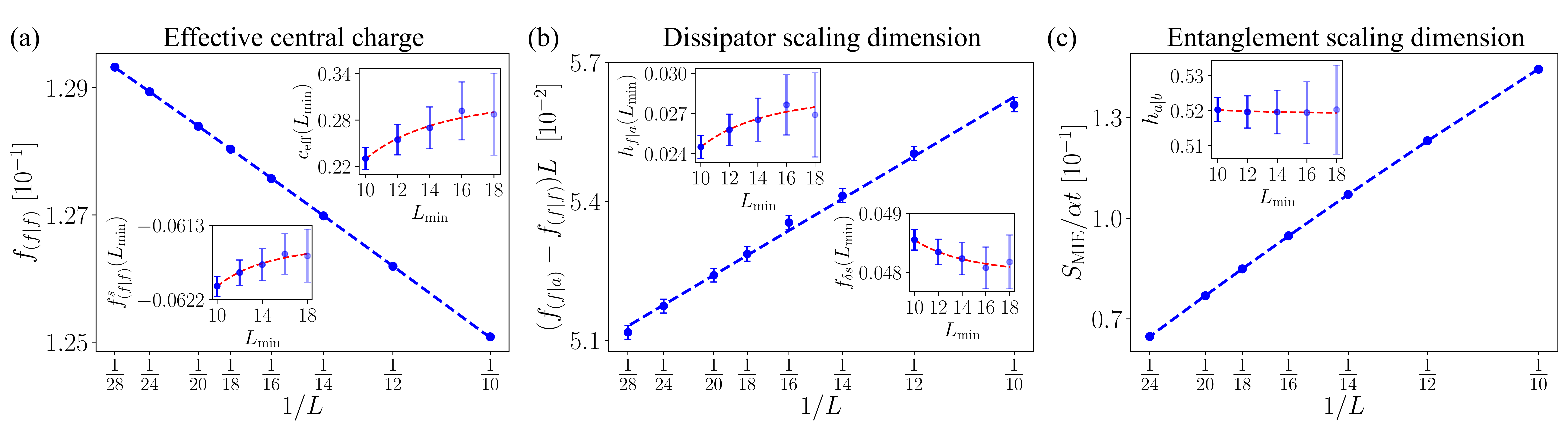}
	\caption{\textbf{Boundary spectrum of dual Clifford circuit:}We implement the boundary condition in Fig~\ref{fig1} to obtain the respective boundary spectrum of dual Clifford circuit. (a) The plot of $f_{(f|f)}$ and $1/L$ clearly respect the scaling form stated in Eq.~\ref{eqn:4} (for $L_{\rm{min}} \in \{10,12,14,16,18\}$) as shown by the dashed blue line. In the inset, we perform the double fitting procedure to extract effective central charge $c_{\rm{eff}}$ and free energy surface term $f_{(f|f)}^{s} $, by successively removing the smallest system size and then fitting the data with $L \geq L_{\rm{min}}$. This gives us $c_{\rm{eff}} = 0.32(3)$ and $f_{(f|f)}^{s} = -0.0615(2)$ as shown by the red dashed line. (b) The plot of $(f_{(f|a)}-f_{(f|f)})L$ scales linearly with $1/L$ as shown by the dashed blue line, and from fitting the value of slope and intercept we obtain the dissipator scaling dimension $h_{f|a} = 0.029(2)$ and $f_{\delta s} = 0.0479$ as shown with their respective red dashed lines in the inset plots. (c) The plot of $S_{\rm{MIE}}$ vs $1/L$ clearly follows the trend stated in Eq.~\ref{eqn:6} and we find the entanglement scaling dimension $h_{a|b} = 0.519(8)$ using the red dashed line in the inset plot. The fit of $S_{\textrm{MIE}}$ include a leading correction term coefficient $A_{1}= -1.833$. All the data shown above for the dual Clifford circuit have $p_{c} = 0.205$, $\alpha = 1$, and $L \in \{10,12,14,16,18,20,24,28\}$ except for $S_{\textrm{MIE}}$ where $L=28$ is absent. We sample $=5\times 10^{6}$ trajectories to find the average value of free energy density.}
	\label{fig4}
\end{figure*}
\subsection{\label{sec:level3c}Entanglement scaling dimension ($h_{a|b}$)}

Finally, the boundary scaling dimension $h_{a|b}$ has been determined both numerically and analytically for the MOIM with cylindrical geometry\cite{PhysRevB.102.094204}, with $h_{a|b} = \left. \ln 4 \frac{d}{dQ}h^{\rm{Potts}}_{1,2}\right|_{Q=1} \simeq 0.096$. This critical exponent corresponds to counting clusters crossing the strip. Here, we extract it using  the boundary ancilla setup of Fig.~\ref{fig1}(c).
We compute the measurement-induced entanglement ($S_{\rm{MIE}}$) between ancilla qubits present on the left and right ends of the boundary where all system qubits after the last time step are measured.  
The result in Fig.~\ref{fig3}(c) agrees with the scaling form Eq.~\ref{eqn:6} and we extract the entanglement scaling dimension $h_{a|b}(L) = 0.097 -\frac{0.06}{L^{2}}$. 
We conclude that $h_{a|b} = 0.097(2)$,  which is in good agreement with the theory prediction. Hence these results completely support our approach to obtain the boundary conformal spectrum. We thus can now generalize it to the more general unitary-measurement dynamics.

\section{\label{sec:level4}Dual Clifford Random Circuit}
The dual Clifford random circuit 
consists of two-qubit dual-unitary Clifford gates and local measurement operators that project along the computational basis states. 

The two-site dual-unitary Clifford gates comprise those Clifford gates which obey the duality relation which makes them 
unitary both in space and time directions~\cite{PhysRevLett.123.210601}. The dual-unitary condition is satisfied by the \text{SWAP} and the \text{iSWAP} 
classes
of the two qubit Clifford operators. This contains a total of 5760 (576 \text{SWAP} + 5184~\text{iSWAP}) gate operations~\cite{barends2014superconducting}. Using these gates fixes the anisotropy parameter $\alpha = 1$, allowing us to extract critical properties more accurately. We then follow the boundary transfer matrix approach outlined in Sec.~\ref{sec:level2}.

\subsection{\label{sec:level4a}Effective central charge ($c_{\rm{eff}}$)}
 We will first start by discussing the calculation of $c_{\rm{eff}}$ in cylindrical geometry (periodic boundary conditions). In Appendix~\ref{sec:levelB} (Fig~\ref{fig8}(b)), we find that the free energy density $f$ shows a clear linear dependence when plotted against $1/L^2$ in accordance with the CFT result stated in Eq.~\ref{eqn:3}. This yields $c_{\rm{eff}} = 0.349(3)$ and $f(L=\infty)= 0.131574(6)$. We now implement dual Clifford circuit with free boundary ends where we find $c_{\rm{eff}} = 0.32(3)$ from the plot between $f_{(f|f)}$ and $1/L$ as shown in fig~\ref{fig4}(a). More precisely, using a double fitting procedure, we find $c_{\rm{eff}}(L) = 0.32- \frac{8.74}{L^2}$ and $f_{(f|f)}^{s}(L) = -0.0615-\frac{0.0569}{L^2}$. The plot respects the scaling form given in Eq.~\ref{eqn:4} and the value of $c_{\rm{eff}}$ is in agreement with the one extracted from periodic boundary conditions. As in the MOIM case, $c_{\rm{eff}}$ has relatively large error bars as a consequence of the free energy surface term. Last, we find that $f_{(f|f)}^{s} = -0.0615(2)$. 

\subsection{\label{sec:level4b}Dissipator scaling dimension ($h_{f|a}$)}
 We now implement boundary dissipation as shown in Fig~\ref{fig1}(b), to extract the dissipator scaling dimension $h_{f|a}$. The plot of $(f_{(f|a)}-f_{(f|f}))L$ vs $\frac{1}{L}$ displays a linear dependence against $1/L$ as shown in Fig~\ref{fig4}(b). Now using the double fitting procedure we find the boundary scaling dimension $h_{f|a}(L) = 0.029 - \frac{0.430}{L^2}$ and the non-universal free energy surface term $f_{\delta s} = 0.0479+\frac{0.0666}{L^2}$. We conclude that $h_{f|a} = 0.029(2)$ and $ f_{\delta s} = 0.0479(3)$. Note that the scaling dimension remains invariant with the type of dissipator used at the boundary, consistent with the fact that both depolarizing and dephasing channels correspond to the same conformally invariant boundary condition $\alpha =a$. 

\subsection{\label{sec:level4c}Entanglement scaling dimension ($h_{a|b}$)} 

The entanglement scaling dimension $h_{a|b}$ is known for Clifford circuit with cylindrical geometry\cite{PhysRevB.98.205136,PhysRevB.100.134306,skinner2019measurement,PhysRevB.104.104305} from the coefficient of the subsytem entanglement entropy.Here, we extract it from the boundary ancilla setup shown in Fig~\ref{fig1}(c). 

We compute the measurement-induced entanglement $S_{\rm{MIE}}$ between ancilla qubits present on the left and right ends of the boundary where all system qubits after the last time step are measured.   The plot in Fig~\ref{fig4}(c) clearly respects the scaling form~\ref{eqn:6} and we extract the entanglement scaling dimension $h_{a|b}(L) = 0.519 +\frac{0.116}{L^{2}}$. The entanglement scaling dimension $h_{a|b} = 0.519(8)$ agrees with existing results, and provides yet another check of our approach. 

While we focused on dual-unitary Clifford circuits in this section, we also checked that we obtain consistent (but less accurate) results for regular Clifford circuits (see Appendix~\ref{sec:levelA}). This confirms the expectation that Clifford and dual-unitary Clifford monitored circuits are in the same universality class.

\section{\label{sec:level5}Dual Haar Random Circuit}

We now come to the case of Haar random circuits undergoing random and local projective measurements.
The qubit chain evolves in time under a bricklayer sequence of discrete timesteps involving two-qubit entangling gates $U_{i,i+1}$ and local $\sigma^z_i$ measurements with probability $p$ as depicted in Fig.~\ref{fig1}. As in the previous section, to avoid the error associated with computing the space time anisotropy, we consider dual-unitary circuits by choosing only the gates $U_{i,i+1}$ that are unitary in space and time such that $\alpha=1$. 
The dual unitary gates are given by, 
\begin{equation}
    U = e^{i \phi}(u_{+} \otimes u_{-})\cdot V[J] \cdot (v_{-} \otimes v_{+}),
\end{equation}
where $\phi, J \in \mathbb{R}$ are chosen randomly from $[0,\pi)$ and $u_{\pm}, v_{\pm} \in \rm{SU(2)}$ are randomly chosen using the Haar measure, and 
\begin{equation}
    V[J] = \exp\left[-i\frac{\pi}{4}\left(\sigma^{x} \otimes \sigma^{x} + \sigma^{y} \otimes \sigma^{y} + J \sigma^{z} \otimes \sigma^{z}\right)\right].
\end{equation}
The statistical self-duality of these models under rotations forces $\alpha = 1$, allowing for more accurate estimate of their critical properties. 
\begin{figure}[t!]
\includegraphics[scale=0.55]{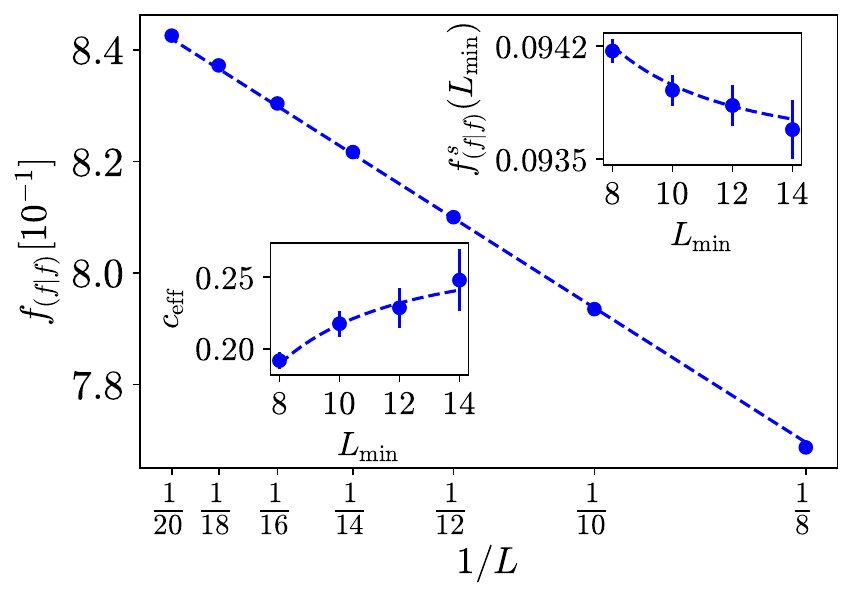}
\caption{{\bf Haar dual unitaries free energy with open boundaries}.  The free energy density $f_{(f|f)}$ with open boundary condition vs $L$. The data fits well with the expected scaling form given in Eq.\eqref{eqn:4} with open boundaries $f_{(f|f)}(L) = f(L=\infty) + 0.0251/L - 0.09417/L^{2}$ allowing us to extract $c_{\rm{eff}}$ and $f^{s}_{(f|f)}$. The top and bottom insets show the double fitting procedure we used to extract $c_{\rm{eff}}$ and $f^{s}_{(f|f)}$ by successively removing the smallest system size($L_{\rm{min}}$) from the fit. For $f^{s}_{(f|f)}$, we find $f^{s}_{(f|f)}(L_{\rm{min}}) = 0.09352 +0.0431/L_{\rm{min}}^{2}$ shown by the dotted line in the top inset.  
For $c_{\rm{eff}}$, we find $c_{\rm{eff}}(L_{\rm{min}}) = 0.27 - 4.92/L^{2}_{\rm{min}}$ shown be the dotted line in the bottom inset. We conclude $c_{\rm{eff}}=0.27(2)$ and $f^s_{(f|f)}=0.09352(2)$. We used $10^7$ trajectories for averaging each data point in this plot.}
\label{fig:HaarObc}
\end{figure}
\subsection{Effective central charge $(c_{\rm{eff}})$}

First, we consider the dynamics of the model with open boundary conditions shown in Fig.~\ref{fig1} (a). 
As discussed in Sec.~\ref{sec:level2},
the effective central charge can be obtained from the free energy density $f_{(f|f)}$ expected to obey the
scaling form given in Eq.~\eqref{eqn:4}.
We note that numerically probing the CFT with OBC is costly as it requires averaging the quantities over a large number of trajectories compared to that with cylindrical geometry. For the results presented in this section, we obtained $10^7$ samples for statistical averaging, similar to the stabilizer simulation presented before.  
In Fig.~\ref{fig:HaarObc}, we show $f_{(f|f)}$ at late times ($t \gg L$) vs $L$ for $L=8$ to $20$. In addition to the leading $1/L$ scaling, $f_{(f|f)}$ exhibits the expected $1/L^2$ correction. As suggested in Eq.\eqref{eqn:4}, the coefficient of $1/L$ term gives $f^s_{(f|f)}$, while that of the $1/L^2$ is related to $c_{\rm{eff}}$.
To obtain estimates of $c_{\rm{eff}}$ and $f^s_{(f|f)}$, we use the double fitting procedure explained previously to fit with the forms, $c_{\rm{eff}}(L)=c_{\rm{eff}}(L=\infty)+ b/L^{2}$ and $f^s_{(f|f)}(L)=f^s_{(f|f)}(L= \infty) + \tilde{b}/L^{2} $ shown in bottom and top insets of Fig.~\ref{fig:HaarObc} respectively. 
 This double fitting procedure gives an estimate of $c_{\rm{eff}} = 0.27(2)$ which matches with the previous result for periodic boundary condition ($c_{\rm{eff}} = 0.24(2)$) obtained in Ref.~\onlinecite{PhysRevLett.128.050602}. We also estimate the surface free energy, $f^s_{(f|f)}=0.09352(2)$. We note that we used the extensive bulk parameter $f(L= \infty)= 0.8902(2)$,  
which is obtained using periodic boundary conditions as we did for the other circuit models.

\begin{figure}[t!]
\includegraphics[scale=0.55]{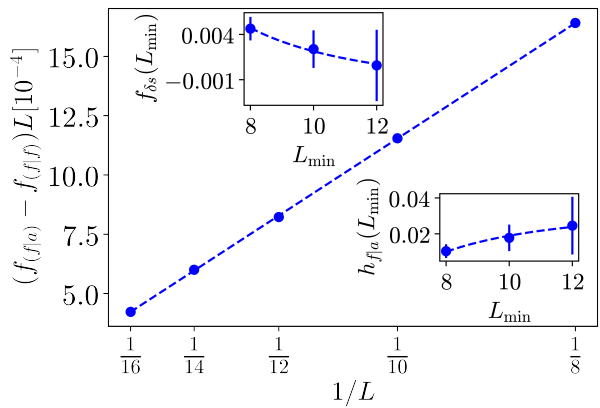}
\caption{{\bf Haar dissipator free energy and its difference with open boundaries}.  The dissipator free energies $f_{\phi}(L)$ displays expected dependence on $L$ from Eq.~\eqref{eqn:4}. For system sizes up to $L=14$ we ran $6 \times 10^8$ trajectories, including Monte Carlo samples. For $L=16$, we ran $6\times 10^7$ trajectories. The difference in free energy densities displays the correct dependence as expected from Eq.~\eqref{eqn:fdiff}, $f_{f|a}(L) - f_{f|f}(L) = \frac{0.004}{L} + \frac{0.077}{L^{2}} $ allowing us to extract $f_{s}$ (Inset top) and $h_{f|a}$ (inset bottom). We find that $f_{\delta s} (L_{\rm min}) = -0.003 + \frac{0.566}{L_{\rm min}^{2}}$ and $h_{f|a}(L_{\rm min}) = 0.05 - \frac{2.26}{L_{\rm min}^{2}}$.} 
\label{fig:Haardiss}
\end{figure}
\begin{table*}[t!]
\caption{\label{tab:table1}The non-universal and universal parameters for various circuit types at critical probability $p_{c}$, the non-universal parameter extracted from the scaling form of free energy density and measurement-induced entanglement are bulk free energy $f(L=\infty)$, open surface free energy $f_{(f|f)}^{s}$, and depolarizer surface free energy $f_{(f|a)}^{s}$. The universal parameters are effective central charge $c_{\rm{eff}}$ with periodic boundary condition (PBC) and open boundary condition (OBC), boundary scaling dimensions; $h_{f|a}$ and $h_{a|b}$.}

\begin{ruledtabular}
\begin{tabular}{ 
|p{2.0cm}||p{1.1cm}|p{1.7cm}|p{1.7cm}|p{1.5cm}||p{1.3cm}|p{1.3cm}|p{1.5cm}|p{1.3cm}| }
 \hline
 \multicolumn{1}{|c||}{} & \multicolumn{4}{c||}{Non-universal parameters} & \multicolumn{4}{c|}{Universal parameters} \\
 \hline
 \multicolumn{1}{|c||}{Circuit type}& $p_{c}$ & $f(L=\infty)$ & $f_{(f|f)}^{s}$ & $f_{(f|a)}^{s}$ & $c_{\rm{eff}}$ (PBC) & $c_{\rm{eff}}$ (OBC) & $h_{f|a}$ & $h_{a|b}$\\
 \hline
 MOIM & 0.5 & 0.55718(2) & -0.4671(7) & -0.1919(7) & 0.96(1) & 1.0(1)  & 0.048(6) & 0.097(2) \\
Dual Clifford & 0.205 & 0.131574(6) & -0.0615(2) & -0.0136(3) & 0.349(3) & 0.32(3)  & 0.029(2) & 0.519(8) \\ 
Clifford & 0.1596 & 0.170469(6)& -0.0749(3)  & -0.0202(3) & 0.375(5)  & 0.32(4) & 0.034(2) & 0.534(2) \\ 
Dual Haar & 0.14 & 0.8902(2) & 0.09352(2) & 0.094(8) & 0.24(2) & 0.27(2) & 0.05(1) & --- \\  
 \hline
\end{tabular}
\end{ruledtabular}
\end{table*}
 
\subsection{Dissipator scaling dimension ($h_{f|a}$)}
We next apply a ``dissipator'' to one end of the open boundary conditions. In particular in the odd time steps, the qubit at the right end of the chain (i.e. site $x=L$) is subjected to the dephasing channel along the z-axis which can be rewritten as, 
\begin{equation}
\mathcal{D}(\rho) =  P_{\uparrow} \rho P_{\uparrow} + P_{\downarrow} \rho P_{\downarrow} = p_d \rho + (1-p_d) \sigma_{z} \rho \sigma_{z},
\end{equation}
where $p_d=\frac{1}{2}$. 
The monitored dynamics with the dissipator can be described by a stochastic master equation, which describes the evolution of the density matrix, $ \dot \rho = \mathcal{L}  \rho$ where the Liouville superoperator $\mathcal{L}$ under the Lindblad approximation takes the form,
\begin{equation}
    \dot \rho = \mathcal{L}  \rho= \hat{c}\rho\hat{c}^{\dagger} - \frac{1}{2} \{\hat{c}\hat{c}^{\dagger},\hat{\rho} \}.
\end{equation}
Here the Lindblad operator $\hat{c}= \sigma_{z}\sqrt{1/(2dt)}$.

Writing the dissipator in this form allows us to Monte Carlo sample many quantum trajectories from a single dephasing channel on the end qubit, and a fixed set of gates and measurement locations on the other qubits. Using this method we can study the dissipative dynamics without having to resort to simulating the full density matrix which is much more numerically challenging(e.g. $\rho$ requires $2^{2L}$ numbers, whereas storing the statevector only requires $2^{L}$. Nonetheless, due to the large number of circuit realizations required to handle the open boundary conditions, how this Monte Carlo sampling is done on top of that, which requires a Monte Carlo average for each circuit sample, remains a non-trivial task. 

As the only quantity we are aiming to compute is the free energy density, we can utilize its dependence on the Monte Carlo sampling ``time'' (denoted by $\tau$ and we stress its not a real physical time) $f_{f|a}(t,L;\tau)$ to extrapolate the Monte Carlo averages to $\tau \rightarrow \infty$. We find that $f_{f|a}(t,L;\tau)$ 
converges like  $1/\tau$ at sufficiently late $\tau$. Instead of converging each circuit sample in Monte Carlo time, we instead work at a fixed Monte Carlo realization of the dissipator, and then average over samples of the circuit. This produces a smooth function that we extrapolate to $\tau \rightarrow \infty$ using a fourth order polynomial in $1/\tau$. Hence, using this $\tau$ dependence, we extract the Monte Carlo average using a reasonable number ($\sim 600$) of Monte Carlo samples. This procedure is demonstrated in more details in Appendix~\ref{sec:MC}. 

With the dissipative free energy in hand $f_{(f|a)}$, we compute its difference with free boundaries
$f_{(f|a)}-f_{(f|f)} = \frac{f_{\delta s}}{L} + \frac{\pi h_{f|a}}{L^2}$ where $f_{\delta s} = f^{s}_{(f|a)}-f^{s}_{(f|f)}$. The results and scaling of the data is shown in Fig.~\ref{fig:Haardiss}.
After performing a similar double fitting procedure, 
we obtain  $f^{s}_{(f|a)} = 0.094(8)$ and find $h_{f|a} = 0.05(1)$. This appears to be different from the dissipator scaling dimension measured in the Clifford case, as expected. We note that this critical exponent is new and was not measured before, and it would be interesting to find other geometries and quantities to extract it. Of course, the boundary-ancilla setup used to measure the entanglement scaling dimension $h_{a|b}$ in the Clifford case cannot be implemented in the Haar case as the number of ancillas grows linearly with time, so our transfer matrix approach cannot be used in that case.

\section{\label{sec:level6}Conclusions}
In this work, we investigated the boundary conformal properties of MIPTs for various circuit models as summarized in Table~\ref{tab:table1} by introducing a numerical boundary transfer matrix approach. This in turn characterizes the CFT that describes these transitions. It further solidifies the fact that Haar and Clifford circuit do not belong to the same universality class. The extracted effective central charges in the cylindrical and infinite strip geometry agree with each other, validating the overall transfer matrix approach as an efficient way to probe MIPTs numerically. This is further validated by the analytically tractable case of the MOIM. In this work, we extracted the scaling dimension associated with two specific 
${\rm BCC}$
operators, namely the dissipator scaling dimension ($h_{f|a}$) and entanglement scaling dimension ($h_{a|b}$). Implementing different boundary conditions
beyond those considered in this paper
should allow one to extract new scaling dimensions, realizing different ``permutations'' of replicas in the statistical mechanics models describing the measurement induced phase transitions discussed in this paper. Classifying these boundary conditions and understanding their physical meaning in terms of quantum information theoretic quantities remains an important challenge for future work.

\begin{acknowledgments}
We thank M. Gullans, D. Huse, A.C. Potter, J. Wilson and A. Zabalo for insightful discussions and for collaboration on previous works. 
This work was partially supported by the Abrahams Postdoctoral Fellowship at the Center for Materials Theory Rutgers (A.C.), the Air Force Office of Scientific Research under Grant No. FA9550-21-1-0123 (A.K., R.V.), the Army Research Office Grant No. 79849-PE-H (K.A., J.H.P.) and a Sloan Research Fellowship (J.H.P., R.V.). 
This work was performed in part  at the Aspen Center for Physics, which is supported by National Science Foundation grant PHY-2210452 (J.H.P.). The authors acknowledge the following research computing resources that have contributed to the results reported here: 
the Open Science Grid~\cite{osg07,osg09}, which is supported by the National Science Foundation award 1148698, and the U.S.\ Department of Energy's Office of Science.
\end{acknowledgments}

\appendix

\begin{figure*}[t!]
	\centering
	\includegraphics[width=\textwidth]{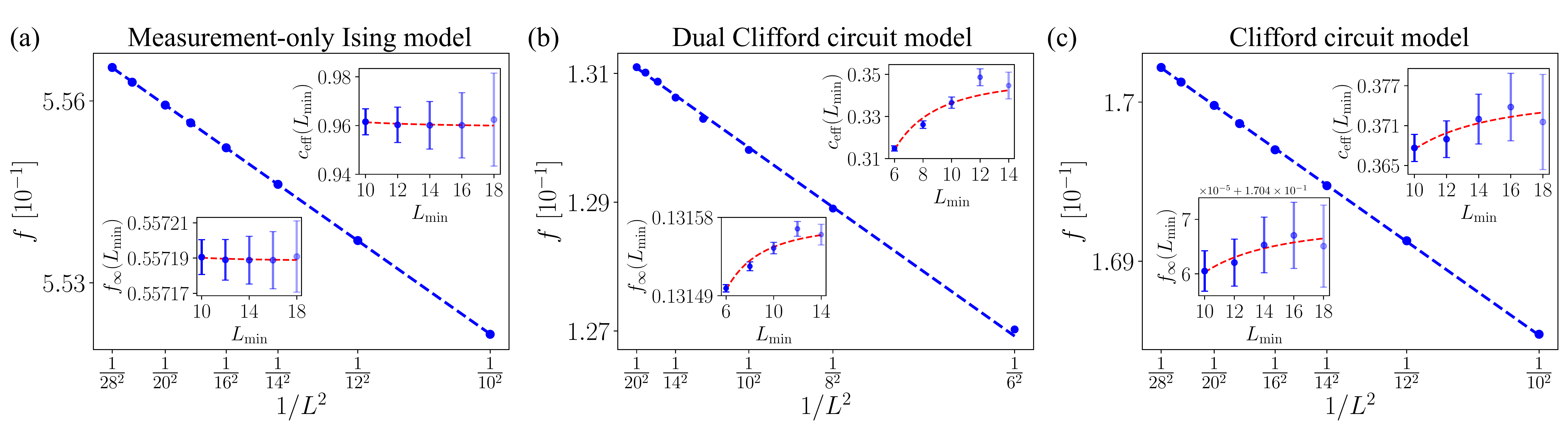}
	\caption{\textbf{Effective central charge in cylindrical geometry:} The plot of $f$ shows linear trend with $1/L^2$ in accordance with the scaling form stated in Eq.~\ref{eqn:3}. We compute $c_{\rm{eff}}$ in cylindrical geometry by the double fitting procedure and we find $c_{\rm{eff}} = 0.96(1),0.349(3),$ and $0.375(5)$ and the extensive bulk parameter $f(L=\infty)= 0.55718(2),0.131574(6),$ and $0.170469(6)$, obtained to high precision for (a)MOIM, (b)dual Clifford, and (c)Clifford random circuits, respectively. Each data point is averaged over $5\times 10^{6}$ trajectories.}
	\label{fig8}
\end{figure*}

\section{\label{sec:levelB} Effective central charge from cylindrical geometry}
The free energy density for cylindrical geometry obeys the relation given by Eq.~\ref{eqn:3}. We plot $f$ vs $1/L^{2}$ for MOIM, dual Clifford, and Clifford random circuits as shown in  Fig~\ref{fig8}. The plot shows a linear trend with $1/L^2$. We extract the value of $c_{\rm{eff}}$ and $f{(\infty)}$ using the double fitting procedure and we find $c_{\rm{eff}} = 0.96(1), 0.349(3), 0.375(5)$, and the bulk term $f(L=\infty) = 0.55718(2), 0.131574(6),$ and $0.170469(6)$ for MOIM, dual Clifford, and Clifford circuit, respectively.

\section{\label{sec:levelC} Boundary dissipation in the measurement-only Ising model}

\begin{figure}[t!]
	\centering
	\includegraphics[width=0.47\textwidth]{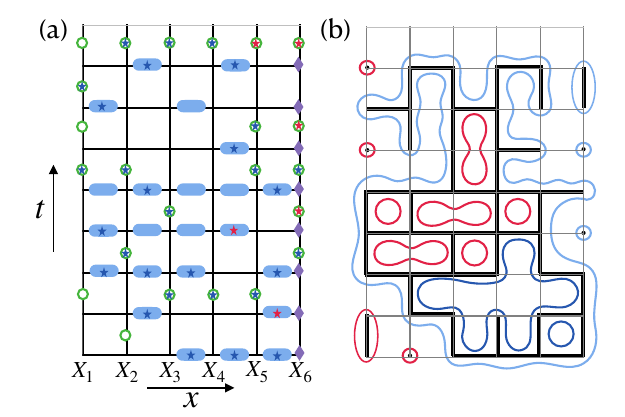}
	\caption{\textbf{Boundary dissipation in the measurement-only Ising model:} (a) The qubits are initialized in a stabilizer state $(X_{1},X_{2},X_{3},X_{4},X_{5}, X_{6})\rm{,~where~} X\equiv \sigma^{x}$, that undergoes measurement plus dissipation dynamics using the two site measurement operation $\sigma^{z}_{i}\sigma^{z}_{i+1}$ (blue edge), the single site measurements $\sigma^{x}_{i}$ (green circles), and the dissipators denoted by purple diamond that act on the right boundary qubit after each time step up to time $t=8$. In this case we consider maximally mixed depolarizing channel to model dissipation. At the last time step, one artificially measures all sites using $\sigma^{x}$ measurement operators. The measurement operations give rise to outcomes that are either random or deterministic. We denote the random outcomes using stars where the red colored stars are the extra random measurements that are absent in the exact same dynamics but with no boundary dissipators. The blue stars are random measurements that are not affected by the boundary dissipation. This particular configuration consists of 6 red and 30 blue stars. (b) Percolation mapping of the configuration shown in (a). The internal and external hulls of boundary clusters (that touch the right boundary) are shown in dark and light blue color, respectively. All other hull of clusters are shown in red. Note that there are 6 boundary hulls, in agreement with eq.~\eqref{eqn:hull}. }
	\label{fig_hull}
\end{figure}

\begin{figure*}[t!]
	\centering
	\includegraphics[width=\textwidth]{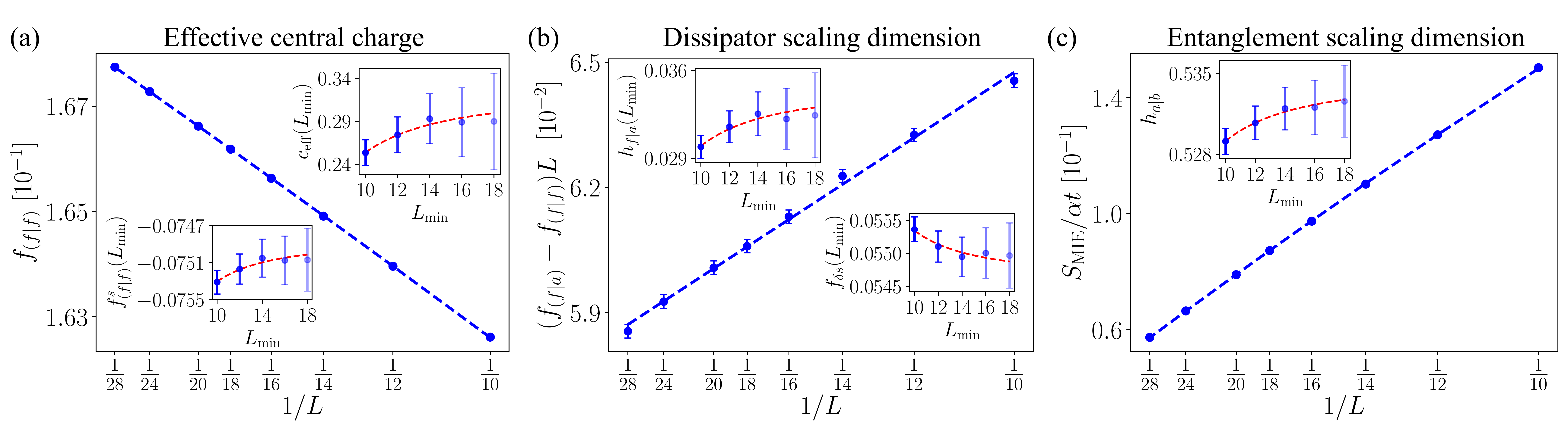}
	\caption{\textbf{Boundary spectrum of Clifford circuit:} We implement the boundary condition in Fig~\ref{fig1} to obtain the respective boundary spectrum of monitored Clifford circuits. (a) The plot of $f_{(f|f)}$ and $1/L$ clearly follows the scaling form Eq.~\ref{eqn:4} (for $L_{\rm{min}} \in \{10,12,14,16,18\}$) as shown by the dashed blue line. In the inset, we perform the double fitting procedure to extract $c_{\rm{eff}}$ and $f_{(f|f)}^{s} $. This gives us $c_{\rm{eff}} = 0.32(4)$ and $f_{(f|f)}^{s} = -0.0749(3)$ as shown by the red dashed line. (b) The plot of $(f_{(f|a)}-f_{(f|f)})L$ scales linearly with $1/L$ as shown by the dashed blue line, and from fitting the value of slope and intercept we obtain $h_{f|a} = 0.034(2)$ and $f_{\delta s} = 0.0547(4)$, shown by the red dashed line in the inset. (c) The plot of $S_{\rm{MIE}}$ vs $1/L$ clearly follows the trend stated in Eq.~\ref{eqn:6} and we find $h_{a|b} = 0.534(2)$ and $A_{1} = -1.9(1)$. All the data shown above for the Clifford circuit have $p_{c} = 0.1596$, $\alpha = 0.61$, and $L \in \{10,12,14,16,18,20,24,28\}$. We sample $=5\times 10^{6}$ trajectories to find the average value of free energy density.}
	\label{fig6}
\end{figure*}
Adding boundary dissipation (depolarizers) to the measurement-only Ising model (MOIM) induces additional non-deterministic, random measurements. In turn, the number of such additional random measurements determines the scaling dimension of the BCC scaling dimension $h_{f|a}$ using eq.~\ref{eqn:fdiff} and 
\begin{equation}
    F_{(f|a)} - F_{(f|f)} = \Delta N_{\rm rand}\ln{2},
    \label{eqn:diff_random}
\end{equation}
where $\Delta N_{\rm rand} = \overline{N}_{\rm rand}^{D} - \overline{N}_{\rm rand}$, with $\overline{N}_{\rm rand}^{D}$ and $\overline{N}_{\rm rand}$ are the averaged number of random measurements with and without the depolarizers present at the right boundary, respectively. We note that for each realization, $\Delta N_{\rm rand}$ has a simple geometrical meaning in terms of percolation clusters. We find by inspection that
\begin{equation}
    \Delta N_{\rm rand}  = N_{\rm bh},
    \label{eqn:hull}
\end{equation}
where $N_{\rm bh}$ represents the number of hulls of clusters touching the right boundary (where the depolarizers act). We illustrate this relation in Fig.~\ref{fig_hull}. This particular example contains $\Delta N_{\rm rand}=6$ extra number of random measurements as compared to the case with no dissipator at the right boundary. Those extra measurements are denoted with red stars in Fig.~\ref{fig_hull}(a). Note that these may happen far from right boundary. The corresponding boundary hulls are shown in Fig.~\ref{fig_hull}(b), there are four ``external'' boundary hulls shown in light blue, and two internal boundary hulls shown in dark blue. The remaining hulls (red) do not contribute to $\Delta N_{\rm rand} $. We have checked the validity of eq.~\eqref{eqn:hull} numerically for very large circuits. 

We now turn to recent boundary CFT results for boundary loop models, that allow us to ``tag'' boundary hulls and count them at criticality~\cite{JACOBSEN2008137}. We find numerically that ``internal'' boundary hulls do not contribute to the universal exponent $h_{f|a}$ (not shown), while the ``external'' boundary hulls can be counted using Ref.~\onlinecite{JACOBSEN2008137}:
\begin{equation}
     \frac{N^{\rm external}_{\rm bh}}{t} \sim \frac{\pi}{2L} \frac{\sqrt{3}}{4 \pi} ,
\end{equation}
ignoring non-universal, ${\cal O} (1)$  contributions, and where the factor of $2L$ is the number of sites in the loop (or Majorana) language. Using eqs.~\eqref{eqn:fdiff} and~\eqref{eqn:diff_random}, we identify the scaling dimension
\begin{equation}
h_{f|a} = \frac{\sqrt{3}}{8 \pi} \ln 2.
\end{equation}

\section{\label{sec:levelA}Clifford random circuits}

 The dynamics of Clifford random circuit can be expressed using the transfer matrix approach discussed in section \ref{sec:level2}. We thus use the scaling form of free energy and $S_{\rm{MIE}}$ to compute the boundary conformal properties for the Clifford MIPT.

\paragraph{\label{sec:levelAa}$c_{\rm{eff}}$.---} 
 We will first start by discussing the calculation of $c_{\rm{eff}}$ in cylindrical geometry. In fig~\ref{fig8}(c) the plot of $f$ shows linear trend with $1/L^2$ in accordance with the CFT result stated in Eq.~\ref{eqn:3}. We note that the double fit method gives $c_{\rm{eff}} = 0.375(5)$ which is in agreement with the previously known result~\cite{PhysRevLett.128.050602} and the extensive bulk parameter $f(L=\infty)= 0.170469(6)$. We now implement Clifford circuit with free boundary ends where we find $c_{\rm{eff}} = 0.32(4)$ from the plot between $f_{(f|f)}$ and $1/L$ as shown in fig~\ref{fig4}(a). The plot respects the scaling form given in Eq.~\ref{eqn:4} and the value of $c_{\rm{eff}}$ validate the result for dual Clifford circuit as noted in Table~\ref{tab:table1}. The extracted values from the plot between $f_{(f|f)}$  and $\frac{1}{L}$ gives leading order correction to Eq.~\ref{eqn:4} since we have $c_{\rm{eff}}(L) = 0.32- \frac{6.608}{L^2}$ and $f_{(f|f)}^{s}(L) = -0.075-\frac{0.042}{L^2}$ using the double fitting procedure.
  
\paragraph{\label{sec:levelAb}$h_{f|a}$.---} 
The plot of $(f_{(f|a)}-f_{(f|f)})L$ vs $\frac{1}{L^2}$ displays a linear trend as shown in \ref{fig6}(b). Now using the double fitting procedure we find the boundary scaling dimension $h_{f|a}(L) = 0.034 - \frac{0.442}{L^2}$ and the non-universal surface term $f_{\delta s} = 0.055+\frac{0.068}{L^2}$.

\paragraph{\label{sec:levelAc}$h_{a|b}$.---} 
The plot in fig~\ref{fig6}(c) clearly respects the scaling form of $S_{\rm{MIE}}$ as given in Eq.~\ref{eqn:6} and we extract the entanglement scaling dimension $h_{a|b}(L) = 0.534 -\frac{0.517}{L^{2}}$ which agrees with known results for Clifford random circuits~\cite{PhysRevB.104.104305}.

\subsection{\label{sec:levelA1}Saturation time}
\begin{figure}[t!]
	\centering
	\includegraphics[width=0.48\textwidth]{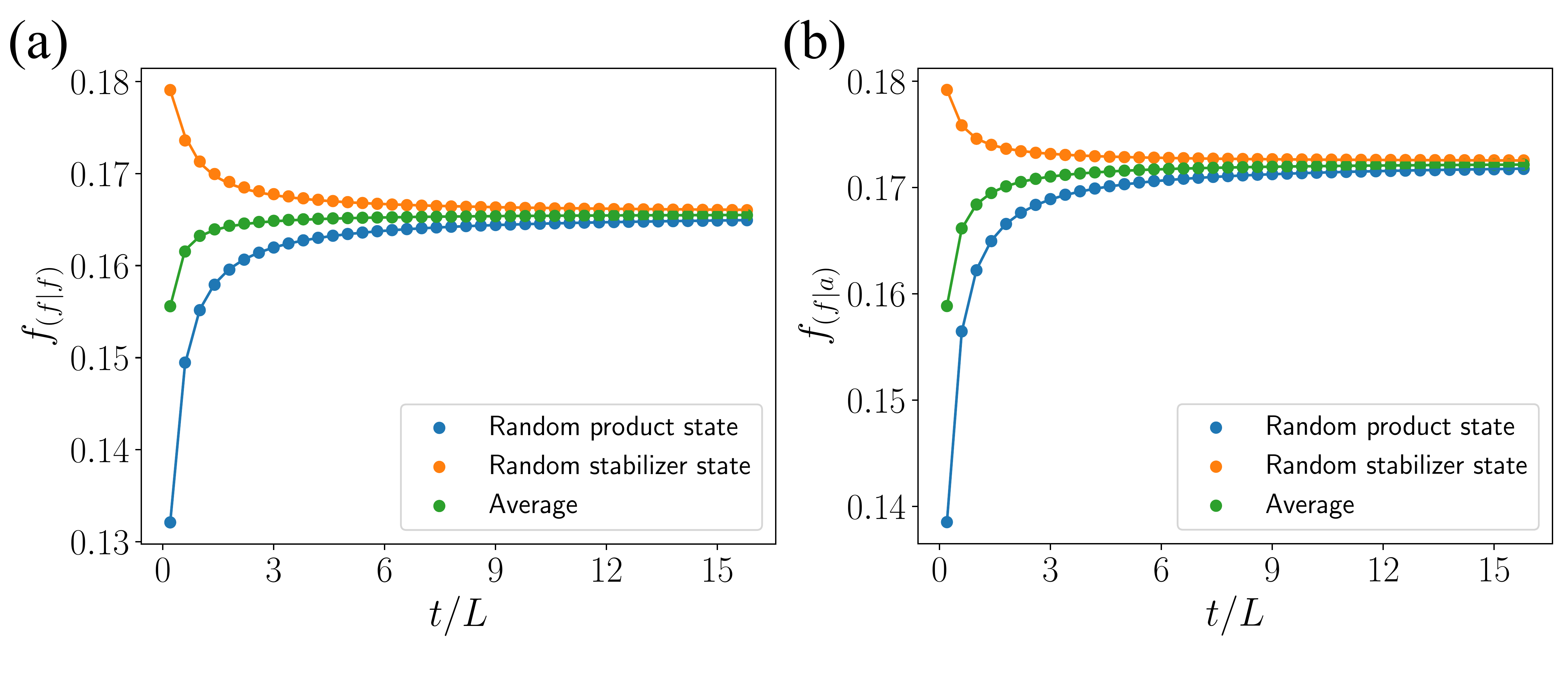}
	\vspace{-2\baselineskip}
	\caption{\textbf{Saturation time:} Evolution of free energy density for Clifford circuit with (a) free boundary ends and (b)with depolarizing channel at the right boundary end. It realizes saturation near the equilibration time $\sim 4L$ both for random product (blue) and random stabilizer initial states(orange). The average of two states together is shown in green.  Following are the parameters for both the plots: $L=16$, $p_{c} = 0.1596$, and number of samples $=10^5$.\vspace{-1\baselineskip}}
	\label{fig7}
\end{figure}

We investigate the dynamics of free energy density to measure the saturation time with and without depolarizers for Clifford random circuit as shown in Fig~\ref{fig7}. We initialize the circuit with a random product or a random stabilizer state and then compute free energy density by averaging over $10^5$ sampled trajectories. One clearly notice saturation above the equilibration time $\sim 4L$ and the average in fig~\ref{fig7} represents the mean of free energy associated with the two types of initial state.

\section {Methods of Monte Carlo Sampling}
\label{sec:MC}
A quantum system undergoing decoherence can be described by the map $\rho(t+dt) = \sum_{k} \hat M_{k} \rho \hat M^{\dagger}_{k}$, where the set of Krauss Operators $\{\hat{M}_{k}\}$ satisfy $\sum_{k} \hat{M}^{\dagger}_{k} \hat{M}_{k} = I$, over the set of possible measurement outcomes $k$. The choice of $M_{k}$ is not necessarily unique. If the state were initially pure, then after a measurement described by $\hat{M}_{k}$, $|\psi \rangle \rightarrow \frac{\hat{M}_{k}|\psi\rangle}{\sqrt{p_{k}}}$, with probability $p_{k} = \mathrm{Tr}({\hat{M}_{k}|\psi\rangle\langle \psi |}\hat{M}^{\dagger}_{k}$).
The Monte Carlo method is implemented by starting from a pure initial state, and evolving $|\psi \rangle$ with $\hat{M}_{k}$ with probability $\langle \psi |M^{\dagger}_{k} M_{k} | \psi \rangle.$
The dephasing channel can be written with $k=2$ Krauss operators: $\hat{M}_{0} = P_{\uparrow}$ and $\hat{M}_{1} = P_{\downarrow}$ 
, where 
\begin{equation}
\rho(t+dt) = \hat M_{0}\rho \hat M_{0}^{\dag} + \hat M_{1} \rho \hat M_{1}^{\dag},
\end{equation}
or alternatively with Krauss operators 
\begin{eqnarray}
    \hat{M}^{'}_{0}(dt) = \mathbb{I} - \frac{c_{1}^{\dag} c_{1}}{2}dt = \sqrt{1-p_d} I, \nonumber \\
    \hat{M}^{'}_{1}(dt) = \sqrt{dt} c_{1}^{\dag} = \sqrt{p_d} \sigma_{z}
\end{eqnarray}
where $\hat{c} = \sqrt{\frac{1-p_d}{dt}}\sigma_{z}$ is the Lindblad operator for a dephasing channel along the z-axis and $p_d=1/2$. In a given Monte Carlo sample of the dissipator, $|\psi \rangle$ undergoes a ``jump'', or evolution with $\hat{c}$, with probability $p_{\mathrm{jump}}= \mathrm{Tr}[\hat{M}^{'}_{1}\rho \hat{M^{'}}^{\dagger}_1]$. In the simulation, a random number $r$, chosen from the unit interval, and if $r < p_{\mathrm{jump}}$, the state vector evolves as $\frac{\hat M'_{1}(dt) \vert \psi(t) \rangle}{\sqrt{\langle \psi(t)|\hat M'^\dag_{1}(dt) \hat M'_{1}(dt)|\psi(t)\rangle}}.$ Whereas, if no jump occurs, $ \vert \psi(t) \rangle$ is unchanged. 

\begin{figure*}[t!]
 \centering
\includegraphics[scale=0.45]{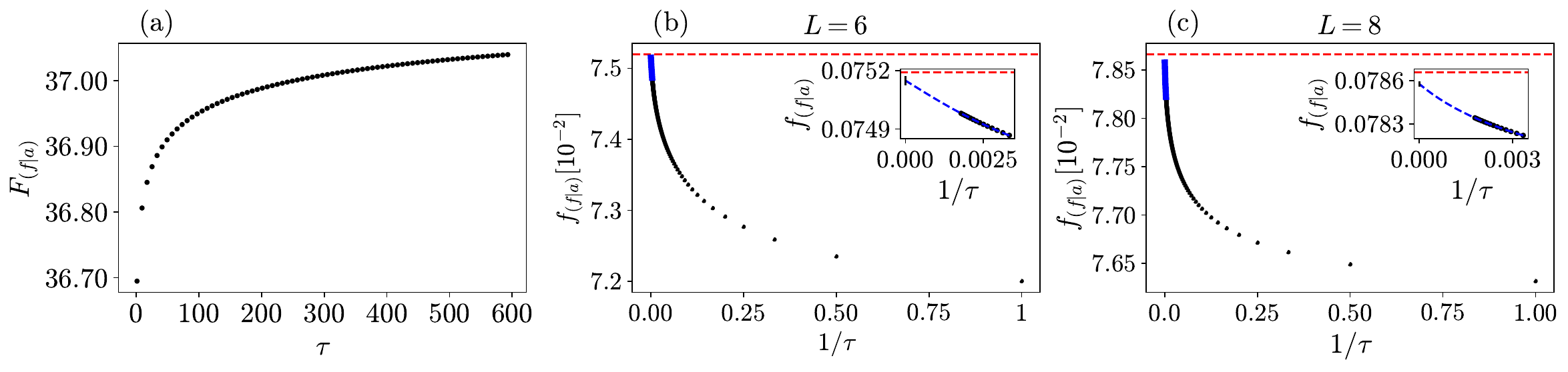}
\caption{ (a) $F_{f|a}$ extrapolated to various values of $\tau$. (b) $f_{f|a}$ as a function of $1/\tau$, where $\tau$ is the number of Monte Carlo iterations used to calculate $f_{f|a}$. The dashed blue line is extrapolated fit, which is also shown in the inset. The dashed red line in the main plot is the full density matrix data. Inset: $f_{f|a}$ computed from $F_{f|a}$ extrapolated to $\tau$. The black line is  computed from the free energy density without extrapolation, starting from $\tau_{it}$ = 300 to $\tau_{it}$ = 600. The blue line is the extrapolation to $\tau_{it} = \infty$  for $L=6$ and $L=8$. For $L=6$, the full density matrix yields $f = \mathrm{0.07519188(1)}$, and $f(\tau=\infty) = 0.07514(2)$ from the extrapolation method. For $L=8$, $f = \mathrm{0.07865890(1)}$, and $f(\tau=\infty) = \mathrm{0.07857(2)}$.}
\label{fig:Haaraggregate}
\end{figure*}

Each circuit realization is specified by a set of two-site random unitary gates $U = \{U_{t,ij}\}$, the space-time measurement locations $\vec{X}$, and the measurement trajectory ${\bf m}$ yielding the outcomes of the measurements. After $T$ timesteps, the circuit yields the unnormalized state: 
\begin{widetext}
\begin{equation}
\rho_{{\bf m}} = \rho_{{\bf m}}(U,\vec{X},T) = \mathcal{D}\{P_{T}U_{T}...\mathcal{D}\{P_{2}U_{2}P_{1}U_{1} \rho U_{1}^{\dag}P_{1}U_{2}^{\dag}P_{2}\}...U^{\dag}_{T}P_{T}\}.
\end{equation}
One timestep consists of a layer of either even or odd gates and measurements. The probability of a meausurement trajectory ${\bf m}$ is $p_{{\bf m}}(U,\vec{X},T) = \mathrm{Tr}(\rho_{{\bf m}}(U,\vec{X},T))$. The dephasing channel can be rewritten as 
\begin{equation}
\rho_{{\bf m}} = \sum_{\vec{i}}^{ } \hat{M}_{i_{T}}\{P_{T}U_{T}...\hat{M}_{i_{2}}\{P_{2}U_{2}P_{1}U_{1} \rho U_{1}^{\dag}P_{1}U_{2}^{\dag}P_{2}\}\hat{M}_{i_{2}}...U^{\dag}_{T}P_{T}\}\hat{M}_{i_{T}},
\end{equation}
where $\hat{M}_{i_{j}}$ is an operator chosen from either Krauss representation and we sum over all $2^{T/2}$ possible dephasing trajectories labelled by $\vec{i} = \{i_{2},...,i_{T}\}$ for fixed $(U,\vec{X},{\bf m})$.
We can now write the probability of a given Lindblad trajectory described by the specific Krauss operators $\mathcal{M}_{i_{2},i_{4},...,i_{T}}=\{M^{'}_{i_{2}},M^{'}_{i_{4}},...,M^{'}_{i_{T}} \}$, where $i_{t} \in \{0,1\}$ labels the Krauss operator for the dephasing channel at the $t_{th}$ layer. For each $(U,\vec{X},{\bf m})$, we apply $\hat{M}^{'}_{0}$ or $\hat{M}^{'}_{1}$ with equal probability on $x=L$ at each timestep. After $T$ timesteps we have, 
\begin{equation}
p_{{i_{2}},i_{4},...,i_{T}} =\mathrm{Tr}\{\hat{M}^{'}_{i_{T}}P_{2T}U_{2T}...\{\hat{M}^{'}_{{i}_{2}}P_{2}U_{2}P_{1}U_{1}\rho U^{\dagger}_{1}P_{1}U^{\dagger}_{2}P_{2} \hat{M}^{'}_{{i}_{2}}\}...U^{\dagger}_{2T}P_{2T} \hat{M}^{'}_{i_{T}}\}.
\end{equation}
We compute $p_{i_{2},i_{4},...,i_{T}}$ for several different trajectories 
and estimate the Born probability after averaging over all possible dissipator outcomes $p_{{\bf m}}$ 
\begin{equation}
p_{{\bf m}}=\left[p_{i_{2},i_{4},...,i_{T}} \right]_{MC}
\end{equation}
where $[\dots]_{MC}$ 
denotes a Monte Carlo average over dissipator outcomes
for many samples. 
\end{widetext}

\subsection{Extrapolation Method}
In this subsection, we describe our numerical method of Monte Carlo sampling from the full density density matrix to approximate the free energy. For a fixed $(U,\vec{X})$, the free energy of the measurement record, $F$, is the average of the logarithm of the probability of a given trajectory: 
\begin{equation}
F = - \sum_{{\bf m}} p_{{\bf m}} \ln p_{{\bf m}} = - \sum_{{\bf m}} \langle \ln p_{{\bf m}} \rangle.
\end{equation}
We perform the Monte Carlo sampling over the dissipator outcomes for a fixed projective measurement trajectory ${\bf m}$ and fixed $(U,\vec{X})$. The probability of a given Monte Carlo trajectory $i$, consisting of a fixed sequence of Monte Carlo propagators from the dissipator, is $p_{{\bf m}}{(i)}.$ Obtaining many samples $i$ allows us to approximate $ p_{{\bf m}} =  [p_{{\bf m}}(i)]_{MC} $ for  a fixed set of gates and measurements. 

The accuracy of the sampling method depends on the number of Monte Carlo samples taken, due to the statistics of the dissipator outcomes. Due to the large number of Monte Carlo trajectories with high $p_{{\bf m}}$, $\langle p_{{\bf m}}(i) \rangle$ decreases as the number of Monte Carlo trajectories increases.  

To estimate the entropy of the measurement record, we record the free energy as a function of the number of Monte Carlo samples (denoted as $\tau$) at each time: $F(t;\tau).$ 
To determine the free energy when $\tau \rightarrow \infty$, we compute the free energy averaged over multiple circuit realizations and outcomes as a function of the number of Monte Carlo iterations. We extrapolate to $\tau \rightarrow \infty$ from $\tau^{\mathrm{min}} = 300$ to $\tau^{\mathrm{max}} = 600$, where $F$ exhibits a leading $1/\tau$  dependence for all system sizes studied, as depicted in Fig.~\ref{fig:Haaraggregate}. We obtain an estimate for $F(t;\tau \rightarrow \infty)$ from the intercept of the extrapolation as $\tau \rightarrow \infty$ using a fourth order polynomial in $1/\tau$.
From the intercept of the extrapolation we estimate $F(t;\tau \rightarrow \infty)$. The accuracy of the approach is exemplified in Fig.~\ref{fig:Haaraggregate} and discussed in more detail in its caption.

To compute the error bars, we record $f$ from individual trajectories, and calculate the bootstrap standard error $\sigma$ for $500$ samples for random Haar and product initial states. The errors are combined using $\sigma = \frac{1}{2}\sqrt{\sigma_{\rm{product}}^{2} + \sigma_{\rm{Haar}}^{2}}$, for each $f(L)$ point. The size of the error bar in $c_{\rm{eff}}$ and $f_{s}$ is determined by the range of possible $c_{\rm{eff}}$ and $f_{s}$ for $f$ varying within $\sigma$.




\bibliography{references}
\end{document}